\documentclass[aps,pra,twocolumn,superscriptaddress]{revtex4-1}
\usepackage{amsmath,amsfonts,amssymb,amscd,mathtools}
\usepackage{dsfont}
\usepackage{amsthm}
\usepackage{graphicx}
\usepackage{dcolumn}
\usepackage{bm}
\usepackage{booktabs}
\usepackage{hyperref}
\usepackage[dvipsnames]{xcolor}
\hypersetup{
    bookmarksnumbered=true, 
    unicode=false, 
    pdfstartview={FitH}, 
    pdftitle={}, 
    pdfauthor={}, 
    pdfsubject={}, 
    pdfcreator={}, 
    pdfproducer={}, 
    pdfkeywords={}, 
    pdfnewwindow=true, 
    colorlinks=true, 
    linkcolor=NavyBlue, 
    citecolor=NavyBlue, 
    filecolor=NavyBlue, 
    urlcolor=NavyBlue 
}
\newcounter{one}
\newcommand{\ketbra}[2]{|{#1}\rangle\!\langle{#2}|}

\newcommand{\Tr}[0]{ \mathrm{Tr}}

\newcommand{\eq}[1]{\begin{align} #1 \end{align}}

\theoremstyle{definition}

\newcommand{\beq}{\begin{equation}}
\newcommand{\eeq}{\end{equation}}

\newcommand{\bal}{\begin{equation}\begin{aligned}}
\newcommand{\eal}{\end{aligned}\end{equation}}

\usepackage[lmargin=.7in,rmargin=.7in,tmargin=.7in,bmargin=1in]{geometry}

\begin{document}
\title{Approaching Carnot Efficiency at Finite Power in an Experimentally Feasible Quantum Heat Engine}
\author{Shogo Toma}
\affiliation{Department of Physics, The University of Tokyo, 7-3-1 Hongo, Bunkyo-ku, Tokyo 113-0033, Japan}
\email{toma-sho5-0611@g.ecc.u-tokyo.ac.jp}
\author{Atsushi Noguchi}
\affiliation{Komaba Institute for Science (KIS), The University of Tokyo, Meguro-ku, Tokyo 153-8902,
Japan}
\affiliation{RIKEN Center for Quantum Computing (RQC), Wako, Saitama 351–0198, Japan}
\affiliation{Inamori Research Institute for Science (InaRIS), Kyoto-shi, Kyoto 600-8411, Japan}
\email{atsushi.noguchi@riken.jp}
\author{Ken Funo}
\affiliation{Department of Applied Physics, The University of Tokyo,  7-3-1 Hongo, Bunkyo-ku, Tokyo 113-8656, Japan}
\email{funo@ap.t.u-tokyo.ac.jp}
\author{Hiroyasu Tajima}
\affiliation{Information Science and Electrical Engineering, Department of Informatics, Kyushu University, Nishi-ku, Fukuoka 819-0395, Japan}
\affiliation{
JST, FOREST, 4-1-8 Honcho, Kawaguchi, Saitama, 332-0012, Japan
			}
\email{hiroyasu.tajima@inf.kyushu-u.ac.jp}

\begin{abstract}
Whether a heat engine can approach Carnot efficiency while maintaining finite power is a fundamental question in finite-time thermodynamics.
For classical Markovian heat engines with local interactions, the power-efficiency trade-off forbids an asymptotic approach to Carnot efficiency at finite power.
In quantum systems, by contrast, degeneracy, symmetry, and collective jumps have been theoretically predicted to enable such an asymptotic attainment by enhancing activity.
It has remained open, however, whether this mechanism can be realized in an experimentally implementable heat engine. 
In this Letter, we propose a superconducting-circuit heat engine that emulates the collective enhancement, thereby enabling an asymptotic approach to Carnot efficiency at finite power.
This result demonstrates that, in an implementable model, such an enhanced dissipative mechanism circumvents the power-efficiency trade-off of classical Markovian engines.
Our work connects abstract bounds in finite-time thermodynamics to a concrete circuit-QED platform and suggests a route toward quantum-device design based on collectively enhanced dissipative processes.
\end{abstract}

\maketitle

\textit{Introduction.---}Efficiency and power are two important factors that characterize the performance of heat engines.
The efficiency $\eta$ of an engine is bounded from above by the Carnot efficiency $\eta_C \equiv 1 - \beta_H/\beta_L$, where $\beta_H$ and $\beta_L$ are the inverse temperatures of the baths~\cite{Carnot1824,Callen1985}.
The upper bound is attained only by a quasistatic Carnot cycle whose power vanishes.
Achieving finite power requires finite-time operation, which drives the engine out of equilibrium and causes entropy production, lowering the efficiency below $\eta_C$.
This tension has long been studied through power-efficiency trade-offs in heat engines~\cite{CurzonAhlborn1975,Esposito2010EMP,Shiraishi2016}.

For classical local Markovian cyclic engines, Carnot efficiency and finite power are incompatible.
Shiraishi, Saito, and Tasaki proved it through the trade-off bound~\cite{Shiraishi2016,Shiraishi2019}
\eq{
  P \leq \beta_L\,\eta(\eta_C-\eta)\,\bar{A},
  \label{eq_SST}
}
where $P$ is the cycle-averaged power and $\bar{A}$ is the cycle-averaged activity, a system-size-dependent quantity that reflects the rate of bath-induced transitions.
Although this type of inequality is valid for both quantum and classical open systems~\cite{Shiraishi2016,Shiraishi2019,TF,FT}, it restricts classical systems more severely than quantum ones:
in classical Markovian engines with local interactions, the activity scales linearly with the system size $N$, $\bar{A} = O(N)$~\cite{Shiraishi2019} and $\eta_C - \eta$ cannot be sent to zero while keeping a finite power, $P = O(N)$. 
The $O(N)$ activity therefore forbids even the asymptotic attainment of Carnot efficiency at finite power.

This no-go result also points to a possible strategy for achieving Carnot efficiency at finite power: a quantum implementation can enhance the activity $\bar{A}$ beyond its usual linear scaling with system size.
For example, if $\bar{A} = O(N^2)$, Eq.~\eqref{eq_SST} allows $\eta_C - \eta = O(1/N)$ and $P = O(N)$ at the same time.
From this perspective, Refs.~\cite{TF,FT} studied thermodynamic speed limits on heat currents in Markovian open quantum systems and showed that quantum coherence within degenerate energy sectors can enhance the scaling of $\bar{A}$, thereby enabling the asymptotic attainment of Carnot efficiency at finite power.

However, it remains an open question whether Carnot performance at finite power is merely a mathematical feature of idealized models, or a physical phenomenon accessible in experiments.
In approaches based on symmetric many-qubit systems, taking the scaling limit requires increasing the number of physical qubits. Moreover, engineering collective jumps or many-body interactions while preserving ideal symmetry imposes stringent experimental constraints.
Indeed, the experimental implementation of quantum heat engines has been proposed and realized in systems such as superconducting circuits~\cite{Uusnakki2025}, trapped ions~\cite{Rossnagel2016,Lindenfels2019,VanHorne2020}, NMR systems~\cite{Peterson2019,Assis2019}, NV centers~\cite{Klatzow2019}, photons~\cite{Kim2022}, and cold atoms~\cite{Bouton2021}, but none of them targets the activity enhancement discussed above.
The experimental realizability of such enhancement mechanisms thus remains unclear.

In this Letter, we propose a superconducting-circuit heat engine that overcomes these obstacles. Its working medium consists of two cavity modes, and it exploits a high-order interaction, the coupler-assisted swap (CAS)~\cite{Shirai2023,Balembois2024,Gao2018,Gao2019,Lescanne2020}, in which $c$ photons are exchanged between the cavities through a driven qubit. The CAS-$c$ interaction takes the form $H_{\mathrm{CAS}}(t) \propto L_c\sigma_+e^{-i\omega_d t} + \mathrm{h.c.}$ with 
\begin{equation}
  L_c = (a_1^\dagger)^c a_2^c \quad (c\in\mathbb{N}),
\end{equation}
and yields $L_c$ as the effective jump operator in the reduced dynamics of the working medium.
This jump operator emulates the activity-enhancement mechanism of Refs.~\cite{TF,FT}~(see the End Matter).
Crucially, previous proposals~\cite{TF,FT} require a distinct collective dissipator for each system size, whose engineering becomes progressively harder as the system size grows.
In our scheme, by contrast, the engine can be scaled up on a fixed circuit by increasing the conserved total photon number $N$ of the working medium, which serves as the effective system size.
For CAS-$2$ its activity indeed scales as $\bar{A} = O(N^2)$, so that the bound [Eq.~\eqref{eq_SST}] permits $\eta_C - \eta = O(1/N)$ at finite power.
Constructing an explicit two-stroke cycle, we demonstrate both analytically and numerically that the desired scaling
\begin{equation}
  \eta = \eta_C - O(1/N), \qquad P = O(N)
  \label{eq:claimed_scaling}
\end{equation}
holds in an experimentally feasible parameter regime.

\begin{figure}
  \centering
  \includegraphics[width=\linewidth]{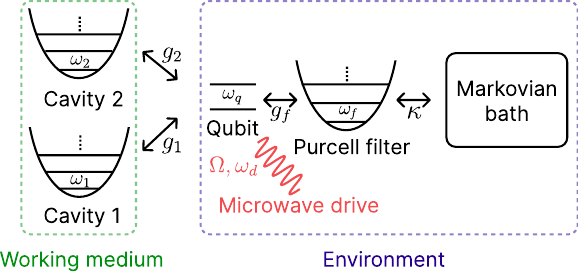}
  \caption{
    Schematic illustration of the experimental setup.
  }
  \label{fig:setup}
\end{figure}

\textit{Experimental setup.---}
We propose a quantum heat engine whose working medium consists of two cavities of frequencies $\omega_1 < \omega_2$, coupled to a common qubit of frequency $\omega_q$~\footnote{Throughout this work, the subscripts $1$ and $2$ refer to cavities~1 and~2, $q$ to the qubit, and $f$ to the Purcell filter.}.
The qubit is in turn coupled to a Purcell-filter cavity~\cite{Purcell1946,Sete2015} of frequency $\omega_f$, which mediates its coupling to a Markovian bath [Fig.~\ref{fig:setup}].
We take $\omega_f$ to be near resonant with the qubit, i.e., $\omega_f \approx \omega_q$.
The full system dynamics is governed by the Gorini–Kossakowski–Sudarshan–Lindblad (GKSL) dynamics~\cite{Gorini1976,Lindblad1976}
\begin{equation}
  \partial_t\rho = -\frac{i}{\hbar}[H(t),\rho] 
    + \kappa(n_\text{th}+1)\,\mathcal{D}[a_f]\rho 
    + \kappa\, n_\text{th}\,\mathcal{D}[a_f^\dagger]\rho,
  \label{eq:full_GKSL_eq}
\end{equation}
with $H(t) = \sum_{j=1,2,f} [\hbar\omega_j a_j^\dagger a_j 
    + \hbar g_j\bigl(a_j^\dagger \sigma_- + a_j \sigma_+\bigr)]+ (\hbar\omega_q/2)\sigma_z 
    + H_\text{drive}(t)$.
Here $\mathcal{D}[L]\rho \equiv L\rho L^\dagger - \tfrac{1}{2}\{L^\dagger L,\rho\}$ 
is the standard Lindblad dissipator, 
$g_{1,2,f}$ denote the qubit--cavity couplings, 
$\kappa$ is the decay rate of the filter cavity, 
$n_\text{th} = (e^{\beta\hbar\omega_f}-1)^{-1}$ is the thermal occupation, 
and $H_\text{drive}(t) = (\hbar\Omega/2)\sigma_+ e^{-i\omega_d t} + \mathrm{h.c.}$ 
is a coherent qubit drive with amplitude $\Omega$ and frequency $\omega_d$.
Throughout this work, we operate in the dispersive regime 
\begin{equation}
\frac{N^{\frac12}g_{1,2}}{\Delta\omega},\, \frac{\Omega}{\Delta\omega} \ll 1,
\label{eq:dispersive regime}
\end{equation}
where $\Delta\omega$ 
denotes the typical magnitude of the relevant detunings~\footnote{More precisely, $\Delta\omega$ denotes the typical magnitude of the detunings appearing in the leading non-resonant processes of the Schrieffer--Wolff expansion.}, so that the 
Schrieffer--Wolff expansion below is perturbatively controlled.
In Eq.~\eqref{eq:full_GKSL_eq} we neglect the intrinsic relaxation and
dephasing of cavities~1,~2 and of the qubit, assuming the engine cycle is
completed on a timescale much shorter than the corresponding coherence
times, so that the engineered dissipation via Purcell filter dominates.

\textit{Effective CAS dissipator.---}
We now derive an effective GKSL equation for the working medium.
We tune the drive frequency to the resonance condition
\begin{equation}
  \omega_d \approx \omega_q - c(\omega_2 - \omega_1),\quad (c\in \mathbb{N}),
  \label{eq: resonance condition of CAS}
\end{equation}
under which a Schrieffer--Wolff transformation~\cite{SW} isolates the 
nearly resonant interaction $H_\mathrm{CAS}$, which coherently transfers $c$ photons between the cavities in conjunction with a qubit transition.
By treating the qubit and the filter cavity coupled to the Markovian bath as an effective environment for the working medium, we derive the reduced dynamics via the Nakajima--Zwanzig projection method~\cite{Nakajima1958,Zwanzig1960}.
With the assumption that the correlation time of this environment is much shorter than the intrinsic timescale of the working medium, we obtain an effective GKSL equation
\begin{align}
  \partial_t \rho_{12} 
  &= -\frac{i}{\hbar}[H_{\mathrm{eff}},\rho_{12}] \notag \\
  &+\gamma_{2\to 1} \mathcal{D}[L_c] \rho_{12} + \gamma_{1\to 2} \mathcal{D}[L_c^\dagger] \rho_{12} \notag \\
  &+ \sum_{j=1,2}\gamma_{\mathrm{loss},j}\Bigl\{ (n_{\mathrm{th}}+1)\mathcal{D}[a_j]\rho_{12}+ n_{\mathrm{th}} \mathcal{D}[a_j^\dagger]\rho_{12}\Bigr\},
  \label{eq:effective_GKSL_eq}
\end{align}
where $\rho_{12} = \mathrm{Tr}_{qf}[\rho]$ is the reduced density matrix of the two cavities, and $\gamma_{2\to 1}$, $\gamma_{1\to 2}$, 
and $\gamma_{\mathrm{loss},j}$ are effective relaxation rates. 
$H_\text{eff} = \hbar \tilde\omega_1 a_1^\dagger a_1 + \hbar \tilde\omega_2 a_2^\dagger a_2 + \cdots$ contains the renormalized cavity frequencies 
$\tilde\omega_i$, together with higher-order corrections such as Kerr and cross-Kerr terms.
The couplings $g_1$ and $g_2$ are chosen so that the dispersive terms proportional to $a_i^\dagger a_i\sigma_z$ cancel.
In the following analysis, we approximate the effective Hamiltonian as $H_\text{eff}\approx\sum_{j=1,2} \hbar \omega_j a_j^\dagger a_j$, since the neglected corrections enter only at fourth order in the perturbative parameters $g_{1,2}/\Delta\omega$ and $\Omega/\Delta\omega$.
See the Supplemental Material Sec.~\ref{sec:Effective_description_of_the_working_medium} for explicit expressions.

In what follows, we further neglect the photon loss terms 
in Eq.~\eqref{eq:effective_GKSL_eq}, obtaining
\begin{equation}
  \partial_t \rho_{12} = -\frac{i}{\hbar} [H_\text{eff}, \rho_{12}] 
    +\gamma_{2\to 1} \mathcal{D}[L_c] \rho_{12} + \gamma_{1\to 2} \mathcal{D}[L_c^\dagger] \rho_{12}.
  \label{eq:coherent_GKSL_eq}
\end{equation}
This approximation, and the derivation above, are justified whenever both the hierarchical regime
\begin{equation}
  \frac{N^{\frac12} g_f^2 g_{1,2}}{\Delta\omega^2} \;\ll\; \Omega\!\left(\frac{N^{\frac12} g_1 g_2}{\Delta\omega^2}\right)^{\!c}
  \;\ll\; \frac{g_f^2}{\kappa},
  \label{eq:approximation_condition}
\end{equation}
and the dispersive regime [Eq.~\eqref{eq:dispersive regime}] hold.
The left inequality ensures $\gamma_\text{loss} \ll \gamma_{2\to 1}$, so that the CAS-$c$ process dominates over photon loss; 
the right inequality enforces the Markovian limit underlying the Nakajima--Zwanzig derivation.
Note that the scaling claimed in this Letter [Eq.~\eqref{eq:claimed_scaling}] is guaranteed only within these two regimes, Eqs.~\eqref{eq:dispersive regime} and \eqref{eq:approximation_condition}.
The derivation of the approximation and further technical details are given in the Supplemental Material Sec.~\ref{sec:Effective_description_of_the_working_medium}.

\textit{Definition of heat and work.---}
Defining heat and work for the working medium is nontrivial here because the coherent drive in the original description [Eq.~\eqref{eq:full_GKSL_eq}] acts on the qubit, which is external to the working medium rather than part of it.
Our strategy is standard: the heat current $J$ is defined so that the second law $\partial_t S_{12} - \beta J \geq 0$ holds, and the extracted work rate $\dot{W}$ is then fixed by 
the first law $\dot{W} = J - \partial_t\langle H_{\mathrm{eff}}\rangle$. Here, $S_{12} \equiv -\mathrm{Tr}[\rho_{12}\ln\rho_{12}]$ is the von Neumann entropy of the working medium.

We define the effective inverse temperature of the working medium $\beta_{\text{eff}}$ by requiring the dissipative part of
Eq.~\eqref{eq:coherent_GKSL_eq} to satisfy detailed balance,
  $\gamma_{2\to 1}/\gamma_{1\to 2} 
  = e^{\beta_{\mathrm{eff}}\, c\hbar(\omega_2-\omega_1)}$.
Since the filter cavity equilibrates with the Markovian bath at the physical inverse temperature 
$\beta$ in the original description Eq.~\eqref{eq:full_GKSL_eq},
$\gamma_{2\to 1}/\gamma_{1\to 2} \approx e^{\beta \hbar\omega_f}\approx e^{\beta \hbar\omega_q}$ holds.
The effective inverse temperature $\beta_{\mathrm{eff}}$ is related to the physical inverse temperature $\beta$ by
$\beta_{\mathrm{eff}} 
  \;\approx\; \beta\,\omega_q/[c(\omega_2-\omega_1)],$
reflecting the fact that the drive converts a bath-induced qubit transition at energy $\omega_q$ into a $c$-photon exchange of energy $c(\omega_2-\omega_1)$ between the cavities.

Together with detailed balance, Spohn's inequality~\cite{Spohn1978} implies the
nonnegativity of the entropy production rate,
  $\dot{\sigma} 
  \;\equiv\; \partial_t S_{12} 
  - \beta_{\mathrm{eff}}\,\mathrm{Tr}\!\left[H_{\mathrm{eff}}\,\partial_t \rho_{12}\right] 
  \;\geq\; 0.$
Therefore, it is natural to define the heat current flowing from the environment into the working medium as
\begin{equation}
  J 
  \;\equiv\; \frac{\beta_{\mathrm{eff}}}{\beta}\,
    \mathrm{Tr}\!\left[H_{\mathrm{eff}}\,\partial_t \rho_{12}\right]
  \;\approx\; \frac{\omega_q}{c(\omega_2-\omega_1)}\,
    \mathrm{Tr}\!\left[H_{\mathrm{eff}}\,\partial_t \rho_{12}\right].
  \label{eq:heat_current}
\end{equation}
The work rate is then fixed by the first law
\begin{equation}
  \dot{W} 
  \;\equiv\; J - \mathrm{Tr}\!\left[H_{\mathrm{eff}}\,\partial_t \rho_{12}\right]
  \;\approx\; \frac{\omega_d}{c(\omega_2-\omega_1)}\,
    \mathrm{Tr}\!\left[H_{\mathrm{eff}}\,\partial_t \rho_{12}\right],
  \label{eq:work_rate}
\end{equation}
where the resonance condition [Eq.~\eqref{eq: resonance condition of CAS}] 
has been used to express the prefactor through the drive frequency $\omega_d$.

Conservation of the total photon number $N = n_1 + n_2$ in the CAS 
dynamics [Eq.~\eqref{eq:coherent_GKSL_eq}] further simplifies these 
expressions. Since $\partial_t\langle n_1\rangle = -\partial_t\langle n_2\rangle$,
we have
$\mathrm{Tr}\!\left[H_{\mathrm{eff}}\,\partial_t \rho_{12}\right] 
  = \hbar(\omega_2-\omega_1)\,\partial_t\langle n_2\rangle$,
so that the heat current and work rate are proportional to the 
photon-transfer rate,
\begin{align}
  J \;\approx\; \frac{\hbar\omega_q}{c}\,\partial_t\langle n_2\rangle,\quad \dot{W} \;\approx\; \frac{\hbar\omega_d}{c}\,\partial_t\langle n_2\rangle.
  \label{eq:JW_compact}
\end{align}
Thus, at each CAS-$c$ event, a quantum of heat $\omega_q$ is absorbed from the bath and a quantum of work $\omega_d$ is extracted through the drive.
In the Supplemental Material Sec.~\ref{sec:heat_and_work_unitary}, we show that these definitions of heat and work 
coincide with those obtained from the underlying unitary dynamics~\cite{Esposito2010}, providing an independent justification 
of Eqs.~\eqref{eq:heat_current} and~\eqref{eq:work_rate}.

\textit{Engine protocol.---}
\begin{figure}
  \centering
  \includegraphics[width=\linewidth]{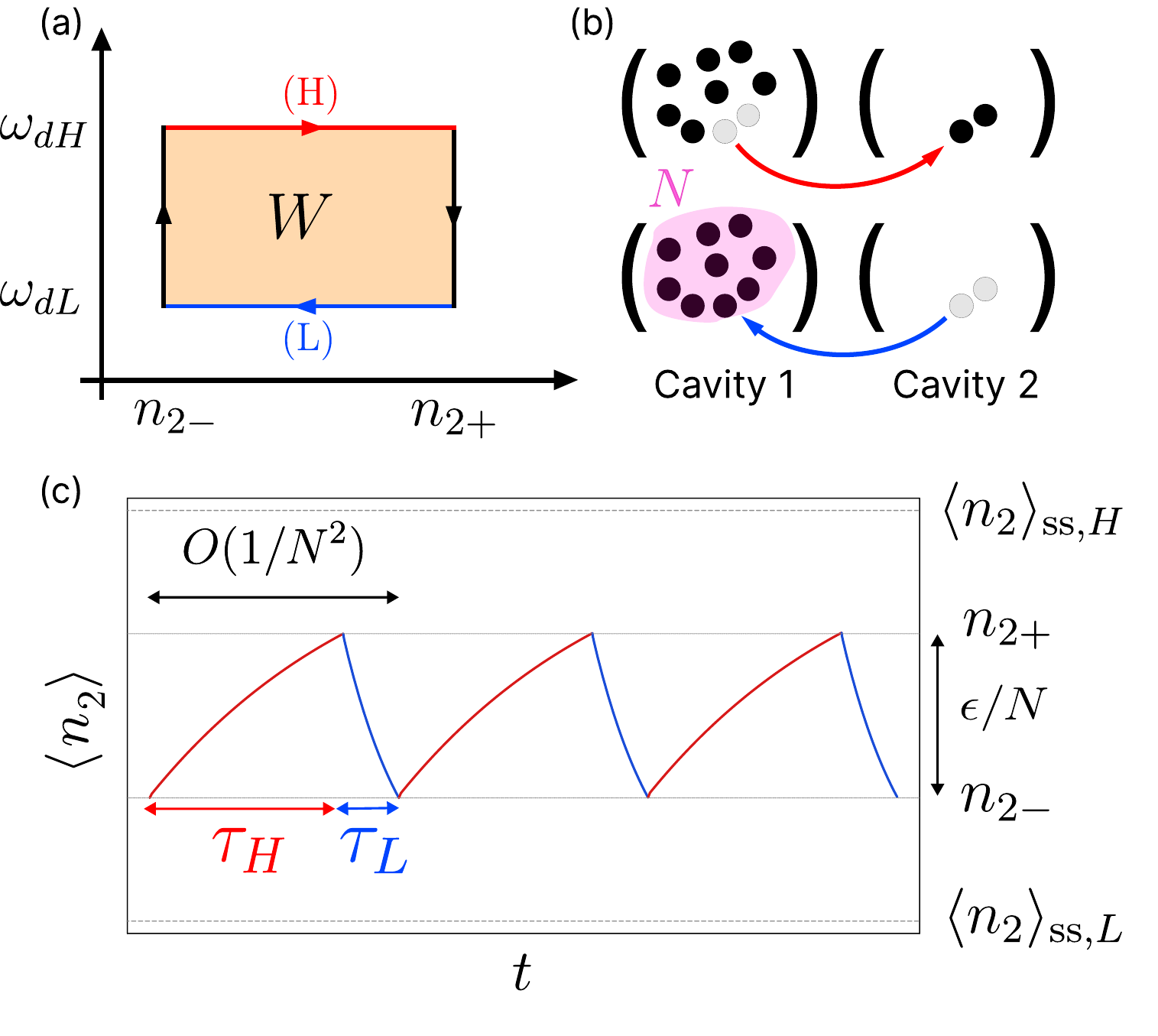}
  \caption{Two-stroke cycle of the CAS-2 engine. 
  (a) Diagram of the cycle in the $(\langle n_2\rangle,\omega_d)$ plane. The area enclosed by the loop gives the work output per cycle, $W=(\hbar/2)(\omega_{dH}-\omega_{dL})\,\Delta\langle n_2\rangle_H$~[cf. Eq.~\eqref{eq:JW_compact}].
  (b) Schematic of the photon exchange realized by the CAS-2 process. Two photons are transferred from cavity 1 to cavity 2 during stroke $\mathrm{(H)}$ (red arrow) and back during stroke $\mathrm{(L)}$ (blue arrow), while the total photon number $N=n_1+n_2$ is conserved.
  (c) Dynamics of $\langle n_2\rangle$ during the first three cycles for $N=2$. Red and blue segments correspond to strokes (H) and (L), respectively.}
  \label{fig:cycle}
\end{figure}
We consider a two-stroke cycle in which the working medium is
alternately coupled to a hot and a cold bath. In each stroke
$\nu=H,L$:
\begin{description}
  \item[($\nu$)] For a duration $\tau_\nu$, the qubit
  and filter cavity are tuned to $\omega_{q\nu}$ and $\omega_{f\nu}$;
  the system is in contact with a Markovian bath at inverse temperature
  $\beta_\nu$ and coherently driven at
  frequency $\omega_{d\nu}$.
\end{description}
The total photon number $N$, fixed by the CAS process,
serves as the scaling parameter of the engine. The frequencies
$\omega_{q\nu}$, $\omega_{f\nu}$, and $\omega_{d\nu}$ are $N$-dependent
and tuned appropriately as the engine is scaled up~\footnote{These $N$-dependent parameters converge to finite values as $N\to\infty$,
because their dependence on $N$ enters only through $\omega_{qL}=(\beta_H/\beta_L)\omega_{qH}+O(1/N)$.}, whereas
$\omega_{1,2}$, $g_{1,2,f}$, and $\Omega$ are kept fixed,
independent of both $N$ and $\nu$.
After sufficiently many repetitions, the cycle approaches a steady cycle in which $\langle n_2\rangle$ and $S_{12}$ return to their initial values once per cycle~\footnote{More precisely, the cycle closes strictly only after the density matrix is projected onto the superselection sector of fixed total photon number $N$ and even parity of $n_2$; see the Supplemental Material Sec.~\ref{sec:numerical_details} for further numerical details.}. 

The thermodynamic quantities of the engine admit simple expressions.
By Eq.~\eqref{eq:JW_compact}, the heat absorbed during stroke~(H) 
and the heat released during stroke~(L) take the compact form
\begin{equation}
  Q_H \;\approx\; \frac{\hbar\omega_{qH}}{c}\,\Delta\langle n_2\rangle_H,
  \qquad
  Q_L \;\approx\; \frac{\hbar\omega_{qL}}{c}\,\Delta\langle n_2\rangle_H,
  \label{eq:QHQL}
\end{equation}
where $\Delta\langle n_2\rangle_H > 0$ is the number of photons 
transferred from cavity~1 to cavity~2 during stroke~(H).
The efficiency and cycle-averaged power then read
\begin{align}
\label{eq:efficiency}
  \eta &\;\equiv\; 1 - \frac{Q_L}{Q_H}\;\approx\; 1 -\frac{\omega_{qL}}{\omega_{qH}},\\
  P &\;\equiv\; \frac{Q_H - Q_L}{\tau_H + \tau_L}\;\approx\; \frac{\hbar(\omega_{qH}-\omega_{qL})}{c}\cdot
    \frac{\Delta\langle n_2\rangle_H}{\tau_H+\tau_L}.
  \label{eq:power}
\end{align}

A quantum extension of Shiraishi--Saito--Tasaki bound [Eq.~\eqref{eq_SST}] is then obtained, with the cycle-averaged activity
\begin{align}
  \overline{A} &= \frac{1}{\tau_H+\tau_L}\int_{\mathrm{cycle}} A(t)\,dt,\notag\\
  A(t) &= \frac{\hbar^2\omega_q(t)^2}{2}\Bigl[\gamma_{2\to1}\langle L_c^\dagger L_c\rangle(t) + \gamma_{1\to2}\langle L_c L_c^\dagger\rangle(t)\Bigr],
\end{align}
where $\langle\,\cdot\,\rangle \equiv \Tr[\,\cdot\,\rho_{12}]$~\cite{TF}.
Near the steady state in which the engine operates, $\overline{A} = O(N^c)$ holds since both $\langle L_c^\dagger L_c\rangle$ and $\langle L_c L_c^\dagger\rangle$ scale as $N^c$.
This enables us to realize $\eta_C - \eta = O(N^{1-c})$ while maintaining finite power.
In particular, CAS-2 admits $\eta = \eta_C - O(1/N)$ together with 
$P = O(N)$, which we show explicitly in the next section.

\textit{Asymptotic realization of Carnot efficiency at finite power.---}
We now exhibit a scaling limit of the CAS-2 engine in which the engine approaches Carnot 
efficiency at finite power. 
We hold $\omega_{qH}$ fixed and tune $\omega_{qL}$ as a function of $N$ such that
  $\langle n_2\rangle_{\mathrm{ss},L} 
  \;=\; \langle n_2\rangle_{\mathrm{ss},H} - \epsilon/N,$
with $\epsilon > 0$ a fixed constant. 
As shown in the Supplemental Material~Sec.~\ref{sec:classical_rate_equation}, $\langle n_2 \rangle$ exhibits approximately exponential relaxation toward
$\langle n_2\rangle_{\mathrm{ss},\nu} \approx 2(e^{\beta_\nu \hbar\omega_{q\nu}}-1)^{-1}$,
with rate
$\Gamma_\nu \equiv 2 N^2 (\gamma_{2\to 1}^\nu + 3\gamma_{1\to 2}^\nu)$.
This leads to
$\beta_L\omega_{qL} - \beta_H\omega_{qH} = O(1/N)$, which,
together with Eq.~\eqref{eq:efficiency}, yields $\eta \;=\; \eta_C - O(1/N)$.
We choose the durations of the strokes (H) and (L) so that, in the steady cycle,
$\langle n_2\rangle$ oscillates between
\begin{align}
  n_{2-} &\equiv (1-\delta)\langle n_2\rangle_{\mathrm{ss},L} 
    + \delta\langle n_2\rangle_{\mathrm{ss},H}, \notag\\
  n_{2+} &\equiv \delta\langle n_2\rangle_{\mathrm{ss},L} 
    + (1-\delta)\langle n_2\rangle_{\mathrm{ss},H},
  \label{eq:endpoints}
\end{align}
where $0 < \delta < 1/2$ is an $N$-independent constant.
That is, stroke (H) drives $\langle n_2\rangle$ from $n_{2-}$ to $n_{2+}$, and stroke 
(L) returns it from $n_{2+}$ to $n_{2-}$ [see Fig.~\ref{fig:cycle}].
This is achieved by choosing the 
stroke durations as
  $\tau_\nu \;=\; \ln[(1-\delta)/\delta]/\Gamma_\nu.$ 
Using Eqs.~\eqref{eq:QHQL} and~\eqref{eq:power}, the cycle-averaged power reads
\begin{equation}
  P \;\approx\; N\frac{\hbar\,(\omega_{qH} - \omega_{qL})(1-2\delta)\epsilon}
             {\ln[(1-\delta)/\delta]\,\mathcal{C}},
  \label{eq:evaluated_power_scaling}
\end{equation}
with
$\mathcal{C} \;\equiv\; 
  (\gamma_{2\to 1}^H+3\gamma_{1\to 2}^H)^{-1}
  + (\gamma_{2\to 1}^L+3\gamma_{1\to 2}^L)^{-1}.$

In the scaling limit $N\to\infty$, both $\omega_{qH} - \omega_{qL} 
\to \omega_{qH} \eta_C = O(1)$ and $\mathcal{C} = O(1)$ 
since $\gamma_{2\to 1}^\nu$ and $\gamma_{1\to 2}^\nu$ remain finite. 
The power therefore asymptotically scales linearly with $N$,
$P = O(N),$
while $\eta$ approaches $\eta_C$ as discussed. 
This is the central result of this Letter: the CAS-2 engine approaches the 
Carnot bound with corrections $O(1/N)$ while sustaining a 
finite power---a regime forbidden to classical Markovian engines.

\textit{Numerical verification of the power scaling.---}
\begin{figure}
  \centering
  \includegraphics[width=\linewidth]{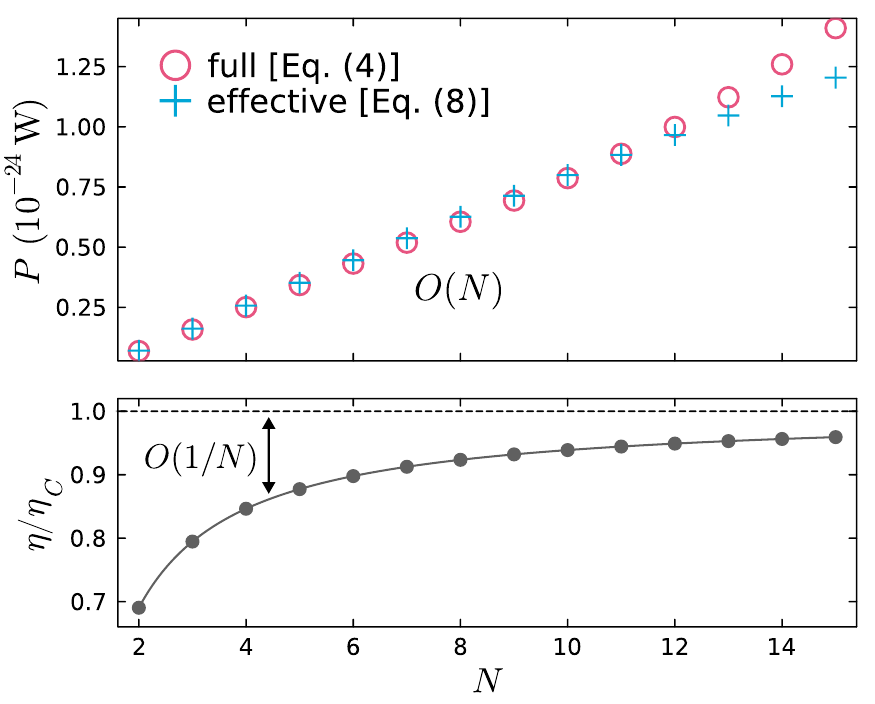}\\
  \caption{Numerical calculations of the CAS-2 engine performance, obtained by directly integrating the full master equation~[Eq.~\eqref{eq:full_GKSL_eq}].
Power and efficiency, evaluated after ten cycles, as functions of the total photon number $N=2,3,\ldots,15$. 
They exhibit the predicted scalings $P=O(N)$ and $\eta_C-\eta=O(1/N)$, respectively.
    The magenta markers show the power obtained from the full master equation [Eq.~\eqref{eq:full_GKSL_eq}], while the light-blue markers show the power obtained from the effective dynamics [Eq.~\eqref{eq:coherent_GKSL_eq}] based on the Schrieffer--Wolff transformation up to fifth order.
    The efficiency $\eta$ is fixed by the choice of $\omega_{q\nu}$ [Eq.~\eqref{eq:efficiency}] and is therefore common to both descriptions.
    All parameters, including $\omega_{q\nu}$, $\omega_{d\nu}$ and $\tau_\nu$,  are optimized to maximize the performance of the full master equation, and the same values are used in both calculations.
    For moderate $N$, the two results agree well, confirming the validity of the effective description.
    See the Supplemental Material Sec.~\ref{sec:numerical_details} for further numerical details.
  }
  \label{fig:performance}
\end{figure}
To assess the validity of this scaling within experimentally accessible 
parameters, we numerically integrate the full master 
equation [Eq.~\eqref{eq:full_GKSL_eq}] under the engine protocol defined in the previous section~\footnote{Since the relaxation dynamics is not purely exponential, we choose $\tau_\nu$ such that the engine cycle operates between $n_{2-}$ and $n_{2+}$.}.
The parameters $(\omega_{1,2}, \omega_{q\nu},\omega_{f\nu}, g_{1,2,f}, \kappa, 
\Omega, \omega_{d\nu})$ are chosen within ranges achievable in 
state-of-the-art superconducting-circuit platforms~\cite{Krantz2019,Tominaga2025,Somoroff2023}.
Figure~\ref{fig:performance} shows the resulting efficiency $\eta$ and 
power $P$ as functions of $N$~\footnote{Parameters: $\omega_1/2\pi =3$~GHz, $\omega_2/2\pi = 3.8$~GHz, $\omega_{qH}/2\pi = 5$~GHz, $\omega_{qL}/2\pi=4.9$~GHz ($N\to \infty$), $g_1/2\pi=50$~MHz, $g_2/2\pi \approx 38$~MHz, $g_f/2\pi=300$~kHz, $\Omega/2\pi=100$~MHz, $\kappa/2\pi = 1$~MHz, $T_H \equiv (k_B\beta_H)^{-1} = 50$~mK, $T_L \equiv (k_B\beta_L)^{-1} = 49$~mK, $\delta=0.3$, $\epsilon = 0.001$.
  All other parameters are optimized so that the power is maximized.}. 
The data confirm the predicted scalings 
$\eta = \eta_C - O(1/N)$ and $P=O(N)$, within the hierarchical window [Eq.~\eqref{eq:approximation_condition}].
This demonstrates that the asymptotic Carnot regime is reached within 
realistic experimental conditions, despite non-resonant processes neglected in the effective description.

\textit{Discussion.---}
We have proposed an experimentally feasible quantum heat engine, based on 
the coupler-assisted swap mechanism in a circuit-QED architecture, that 
asymptotically approaches Carnot efficiency at finite power. 
The collective enhancement of the activity, which scales as 
$N^c$ for the CAS-$c$ process, evades the Shiraishi--Saito--Tasaki trade-off 
forbidden to classical Markovian engines. 
For CAS-2, we showed that an explicit scaling limit realizes 
$\eta = \eta_C - O(1/N)$ together with $P = O(N)$, 
and verified this prediction by direct numerical simulation of the full 
master equation~[Eq.~\eqref{eq:full_GKSL_eq}] using parameters within reach of current superconducting-circuit experiments.

Our results suggest several possible directions for improving the performance of heat engines, as well as other nonequilibrium devices. First, although our numerical demonstration focused mainly on CAS-2 in order to remain within an experimentally realistic parameter regime, there is no fundamental reason to stop at the second-order CAS process. In principle, higher-order CAS processes can generate stronger collective enhancement of the activity, $\bar{A}=O(N^c)$ with $c>2$. This opens the possibility of realizing heat engines that approach Carnot efficiency while maintaining power that scales quadratically, or even superquadratically, with the number of constituents, as predicted in Ref.~\cite{FT}. Establishing concrete circuit-QED implementations of such higher-order CAS processes is an important future direction.

Second, the applicability of the CAS mechanism is not limited to heat engines. The essential ingredient of our construction is the simulation of collective jump operators in an effective many-qubit symmetric subspace using controllable photonic modes. Therefore, the same idea may be useful for a broader class of tasks whose performance can be enhanced by collective dissipation or collective transitions. Possible examples include quantum batteries, quantum sensing, and other nonequilibrium protocols in which the relevant speed, power, or precision is constrained by activity. 
We leave these extensions to future work.

\textit{Acknowledgments.---}
ST was supported by FoPM, WINGS Program, the University of Tokyo.
KF was supported by JST ERATO Grant No. JPMJER2302, Japan, and JSPS KAKENHI Grant Nos. JP23K13036 and JP26H02012.
HT was supported by JSPS Grants-in-Aid for Scientific Research 
No. JP25K00924, and MEXT KAKENHI Grant-in-Aid for Transformative
Research Areas B ``Quantum Energy Innovation'' Grant Numbers 24H00830 and 24H00831, JST MOONSHOT No. JPMJMS2061, and JST FOREST No. JPMJFR2365. 

\textit{Data availability.---}
The data that support the findings of this article are not publicly available. The data are available from the authors upon reasonable request.

\bibliography{ref}





\onecolumngrid
\begin{center}
  \textbf{\large End Matter}
\end{center}
\twocolumngrid

\textit{CAS as superradiance on the Dicke ladder.---}
The $O(N^c)$ scaling of the activity has a transparent
origin in a hidden superradiance structure of the CAS dynamics.
The Schwinger-boson representation
\begin{equation}
  S_- = a_1^\dagger a_2,\qquad S_z = (n_2 - n_1)/2,
\end{equation}
maps the fixed-$N$ two-mode Hilbert space onto the fully symmetric
subspace of $N$ spin-$1/2$'s with total spin $s = N/2$---the Dicke
ladder. Under this mapping, $L_c = (a_1^\dagger)^c a_2^c \mapsto (S_-)^c$,
and Eq.~\eqref{eq:coherent_GKSL_eq} is unitarily equivalent to
\begin{equation}
  \partial_t\rho 
  = -\frac{i}{\hbar}\!\left[\hbar(\omega_2-\omega_1)\,S_z,\rho\right]+ \gamma_{2\to 1}\mathcal{D}[(S_-)^c]\rho 
  + \gamma_{1\to 2}\mathcal{D}[(S_+)^c]\rho.
  \label{eq:dicke}
\end{equation}
The activity on the Dicke ladder,
$A \propto \gamma_{2\to 1}\langle (S_+)^c(S_-)^c\rangle 
+ \gamma_{1\to 2}\langle (S_-)^c(S_+)^c\rangle$,
scales as $O(N^c)$ near thermal equilibrium, reflecting
the $O(N^c)$
collective transition matrix elements within the symmetric subspace.
For $c=1$, Eq.~\eqref{eq:dicke} is precisely the Dicke superradiance 
model~\cite{Dicke1954}, with activity $O(N)$. For $c\geq 2$, the scaling becomes $O(N^c)$,
but $(S_\pm)^c$ are $c$-body nonlocal operators whose direct implementation as dissipators is generally infeasible. The CAS architecture
circumvents this obstruction: the same collective channel is realized
natively by a single $c$-photon transition between two cavity modes without assembling $N$ physical qubits or engineering nonlocal
dissipators.



\newcommand{\beginsupplement}{
  \setcounter{equation}{0}
  \renewcommand{\theequation}{S\arabic{equation}}%
  \renewcommand{\theHequation}{S\arabic{equation}}%
  \setcounter{figure}{0}
  \renewcommand{\thefigure}{S\arabic{figure}}%
  \renewcommand{\theHfigure}{S\arabic{figure}}%
  \setcounter{table}{0}
  \renewcommand{\thetable}{S\arabic{table}}%
  \renewcommand{\theHtable}{S\arabic{table}}%
  \setcounter{section}{0}
  \setcounter{secnumdepth}{2}
}


\onecolumngrid
\clearpage
\begin{center}
  \textbf{\large Supplemental Material}
\end{center}
\beginsupplement
\section{Review: the activity bound and the power--efficiency trade-off}
\label{sec:framework}
In this section, we summarize the general framework underlying the Shiraishi--Saito--Tasaki (SST) trade-off~\cite{Shiraishi2016,Shiraishi2019} and its quantum extension by Tajima and Funo~\cite{TF,FT}, which together motivate the CAS mechanism of the main text. The entire discussion reduces to the classical case when the state is diagonal in the energy eigenbasis.

\subsection{Setting and thermodynamic quantities}
We consider a system coupled to two baths at inverse temperatures $\beta_\nu$ ($\nu = H,L$), evolving under the GKSL equation
\begin{equation}
  \partial_t\rho = -\frac{i}{\hbar}[H(t),\rho] + \sum_\nu\mathcal{D}_\nu\rho,
  \qquad
  \mathcal{D}_\nu\rho = \sum_{\omega,k}\gamma_{\nu,\omega,k}\,\mathcal{D}[L_{\nu,\omega,k}]\rho,
  \label{eq:framework_GKSL}
\end{equation}
where $\nu$ labels the bath, $\omega$ the energy change of a jump, and $k$ distinguishes processes with the same $\nu$ and $\omega$. The jump operators are eigenoperators of the Hamiltonian, $[H,L_{\nu,\omega,k}] = -\omega L_{\nu,\omega,k}$, come in adjoint pairs $L_{\nu,-\omega,k} = L_{\nu,\omega,k}^\dagger$, and satisfy local detailed balance,
\begin{equation}
  \frac{\gamma_{\nu,-\omega,k}}{\gamma_{\nu,\omega,k}} = e^{-\beta_\nu\hbar\omega}.
  \label{eq:framework_detailed_balance}
\end{equation}
These conditions ensure that $L_{\nu,\omega,k}^\dagger L_{\nu,\omega,k}$ is energy-diagonal and that the Gibbs state $\rho^{\beta_\nu}\propto e^{-\beta_\nu H}$ is a stationary state of $\mathcal{D}_\nu$.

The heat current flowing from bath $\nu$ into the system, the extracted work rate, and the entropy production rate are
\begin{equation}
  J_\nu \equiv \Tr[H\,\mathcal{D}_\nu\rho],
  \qquad
  \dot{W} \equiv -\Tr[(\partial_t H)\rho],
  \qquad
  \dot{\sigma} \equiv \partial_t S - \sum_\nu\beta_\nu J_\nu,
  \label{eq:framework_JWsigma}
\end{equation}
with $S = -\Tr[\rho\ln\rho]$. They satisfy the first law $\partial_t\langle H\rangle = \sum_\nu J_\nu - \dot{W}$ and the second law $\dot{\sigma}\geq 0$. The latter follows by writing $\dot{\sigma} = -\sum_\nu\Tr[\mathcal{D}_\nu\rho\,(\ln\rho - \ln\rho^{\beta_\nu})]$ and applying the data-processing inequality to the completely positive map generated by $\mathcal{D}_\nu$, which fixes $\rho^{\beta_\nu}$. A heat cycle is a periodic evolution over $[0,\tau]$ with $\Delta S = \Delta\langle H\rangle = 0$; its efficiency and cycle-averaged power are
\begin{equation}
  \eta \equiv 1 - \frac{Q_L}{Q_H},
  \qquad
  P \equiv \frac{Q_H - Q_L}{\tau},
\end{equation}
where $Q_H$ is the heat absorbed during stroke~(H) and $Q_L$ the heat released during stroke~(L).

\subsection{The activity bound and the SST trade-off}
The key quantity controlling the trade-off is the \emph{activity}
\begin{equation}
  A(t) \equiv \frac{1}{2}\sum_{\nu,\omega,k}(\hbar\omega)^2\,\gamma_{\nu,\omega,k}\,\Tr[L_{\nu,\omega,k}^\dagger L_{\nu,\omega,k}\,\rho(t)],
  \label{eq:framework_activity}
\end{equation}
a purely kinetic measure of how fast the dissipative jumps occur, weighted by the squared energy exchanged. 
Under the above assumptions, Ref.~\cite{TF} shows the following time-local bound:
\begin{equation}
  \frac{J(t)^2}{\dot{\sigma}(t)} \leq A(t),
  \qquad J(t) \equiv |J_H(t)| + |J_L(t)|.
  \label{eq:framework_activity_bound}
\end{equation}
Integrating Eq.~\eqref{eq:framework_activity_bound} over a cycle, again with the Cauchy--Schwarz inequality and the steady-cycle identity $\int_{\mathrm{cycle}}\dot\sigma\,dt = \beta_L Q_H(\eta_C - \eta)$, gives the SST trade-off
\begin{equation}
  P \leq \beta_L\,\bar{A}\,\eta(\eta_C - \eta),
  \qquad
  \bar{A} \equiv \frac{1}{\tau}\int_{\mathrm{cycle}}\,A(t) dt,
  \label{eq:framework_SST}
\end{equation}
which reproduces Eq.~\eqref{eq_SST}. Since $A$ is bounded whenever all system parameters are finite, Eq.~\eqref{eq:framework_SST} forbids $\eta\to\eta_C$ at nonzero power unless $P/\bar{A}\to 0$, that is, unless the activity grows faster than the power. Running $N$ independent copies of an engine makes both quantities extensive, $P = O(N)$ and $\bar{A} = O(N)$, so $P/\bar{A} = O(1)$ and the Carnot efficiency cannot be attained. The asymptotic approach to Carnot efficiency at finite power therefore hinges entirely on the relative scaling of $P$ and $\bar{A}$ with system size.

\section{Effective description of the working medium}
\label{sec:Effective_description_of_the_working_medium}
In this section, we derive the effective CAS dynamics [Eq.~\eqref{eq:coherent_GKSL_eq}] of the working medium from the full master equation [Eq.~\eqref{eq:full_GKSL_eq}] in two steps. First, a time-dependent Schrieffer--Wolff transformation isolates the slowly varying dynamics, yielding a near-static Hamiltonian $H_{\mathrm{SWT}}$ that includes $H_{\mathrm{CAS}}$, together with a transformed filter dissipator.
Second, treating the qubit and the Purcell filter as an effective environment, a Nakajima--Zwanzig projection followed by the Born--Markov approximation eliminates these degrees of freedom and produces a time-local GKSL equation for the two cavities, in which the bath-mediated qubit transitions appear as the collective CAS jumps $\mathcal{D}[L_c]$ and $\mathcal{D}[L_c^\dagger]$.
We close by collecting the assumptions behind the derivation and showing that they reduce to the hierarchical window [Eq.~\eqref{eq:approximation_condition}] of the main text.

\subsection{General framework for the Schrieffer--Wolff transformation}
To isolate the long-time dynamics of Eq.~\eqref{eq:full_GKSL_eq}, we first apply the time-dependent unitary transformation
\begin{equation}
  \label{eq:def_of_tilde_rho}
  \tilde{\rho}^I = e^{S^I(t)}\,\rho^{I}\,e^{-S^I(t)}.
\end{equation}
Here, $X^{I}$ denotes the interaction picture of an operator $X$ with respect to $H_0 = \sum_{i=1,2,f}\hbar\omega_i a_i^\dagger a_i + \frac{\hbar\omega_q}{2}\sigma_z$, and $S(t)$ is a time-dependent anti-Hermitian operator to be determined below. Under this transformation, Eq.~\eqref{eq:full_GKSL_eq} becomes
\begin{equation}
  \partial_t\tilde{\rho}^I = -\frac{i}{\hbar}[H_{\mathrm{SWT}}^I(t),\tilde{\rho}^I]
    + \kappa(n_\text{th}+1)\,\mathcal{D}[\tilde{a}_f^I(t)]\tilde{\rho}^I
    + \kappa\, n_\text{th}\,\mathcal{D}[(\tilde{a}_f^I)^\dagger(t)]\tilde{\rho}^I,
    \label{eq:GKSL_eq_for_tilde_rho}
\end{equation}
with $H_{\mathrm{SWT}}^I(t)= e^{S^I(t)}H^{I}(t)e^{-S^I(t)} + i\hbar\,\partial_t(e^{S^I(t)})\,e^{-S^I(t)}$ and $\tilde{a}_f^I(t) = e^{-i\omega_f t}\,e^{S^I(t)}a_f e^{-S^I(t)}$.

The derivative of the exponential is given by the Wilcox formula~\cite{Wilcox1967},
\begin{equation}
  \partial_t(e^{S^I(t)})\,e^{-S^I(t)} = \int_0^1 \! du\, e^{S^I(t)u}\,\dot{S}^I(t)\,e^{-S^I(t)u}
    = \sum_{n=0}^\infty \frac{1}{(n+1)!}\,(\operatorname{ad}_{S^I})^n(\dot{S}^I(t)),
\end{equation}
where $\operatorname{ad}_A$ denotes the adjoint action, $\operatorname{ad}_A(B) = [A,B]$.
Throughout the Supplemental Material, a dot denotes a time derivative, e.g., $\dot{X}(t)\equiv \partial_t X(t)$.
The first term of $\tilde H^I(t)$ can likewise be expanded and we obtain
\begin{equation}
  H_{\mathrm{SWT}}^I(t) = \sum_{n=0}^\infty \frac{1}{n!}\,(\operatorname{ad}_{S^I})^n\!\left(H^{I}(t) + \frac{i\hbar\,\dot{S}^I(t)}{n+1}\right)\equiv\sum_{n=1}^\infty H_{\mathrm{SWT}}^{I(n)}(t).
\end{equation}
Here, $H_{\mathrm{SWT}}^{I(n)}$ denotes the $O(\lambda^n)$ contribution to $H_{\mathrm{SWT}}^I$.

As noted in the main text, we work in the dispersive regime $N^{\frac12} g_{1,2}/\Delta\omega,\ \Omega/\Delta\omega \ll 1$, where $\Delta\omega$ is the typical detuning, so that the Schrieffer--Wolff expansion below is perturbatively controlled.

Introducing a formal parameter $\lambda$ that tracks the order of magnitude of $g_{1,2}$, and $\Omega$, so that the interaction-picture coupling obeys $H^I = \mathcal{O}(\lambda)$, we expand
\begin{equation}
  S^I(t) = \sum_{n=1}^{\infty} \lambda^n S^I_n(t),
\end{equation}
and fix $S_n$ order by order in $\lambda$, starting from the lowest order, so as to remove the oscillating terms from $H_{\mathrm{SWT}}^I$. 
To make the procedure transparent, we evaluate $S_n$ explicitly at the lowest few orders.

\textit{First order, $\mathcal{O}(\lambda)$.---}Retaining only the $\mathcal{O}(\lambda)$ terms in $H_{\mathrm{SWT}}^I$, we have
\begin{equation}
  H_{\mathrm{SWT}}^{I(1)} = i\hbar\dot{S}_1^I + H^I.
\end{equation}
We assign the slowly varying part of $H^I$ to $H_{\mathrm{SWT}}^{I(1)}$ and let the remaining rapidly oscillating terms be carried by the $i\hbar\dot{S}_1$ contribution. This is achieved by choosing the generator
\begin{equation}
  S_1^I \equiv \frac{i}{\hbar}\int^t \! du\, \bigl(H^I(u) - H_{\mathrm{SWT}}^{I(1)}(u)\bigr).
\end{equation}
Here $\int^t$ denotes an indefinite integral; in what follows we fix the integration constants so that each $S_n$ consists of purely oscillating terms of the form $\mathcal{N}\,\dfrac{g^{k}\Omega^{l}}{(\Delta\omega)^{m}}\,e^{i\Delta\omega t}$, with $\mathcal{N}$ a numerical constant and no secular (zero-frequency) component.

\textit{Second order, $\mathcal{O}(\lambda^2)$.---}Proceeding in the same way,
\begin{equation}
  H_{\mathrm{SWT}}^{I(2)} = i\hbar\dot{S}_2^I + K_2,
  \qquad
  S_2^I \equiv \frac{i}{\hbar}\int^t \! du\, \bigl(K_2(u) - H_{\mathrm{SWT}}^{I(2)}(u)\bigr),
\end{equation}
with
\begin{equation}
  K_2 \equiv \Bigl[S_1^I,\, \tfrac{i\hbar}{2}\dot{S}_1^I + H^I\Bigr].
\end{equation}

\textit{Third order, $\mathcal{O}(\lambda^3)$.---}Likewise,
\begin{equation}
  H_{\mathrm{SWT}}^{I(3)} = i\hbar\dot{S}_3^I + K_3,
  \qquad
  S_3^I \equiv \frac{i}{\hbar}\int^t \! du\, \bigl(K_3(u) - H_{\mathrm{SWT}}^{I(3)}(u)\bigr),
\end{equation}
with
\begin{equation}
  K_3 \equiv \Bigl[S_1^I,\, \tfrac{i\hbar}{2}\dot{S}_2^I\Bigr]
   + \Bigl[S_2^I,\, \tfrac{i\hbar}{2}\dot{S}_1^I + H^I\Bigr]
   + \tfrac{1}{2}\Bigl[S_1^I,\, \bigl[S_1^I,\, \tfrac{i\hbar}{3}\dot{S}_1^I + H^I\bigr]\bigr].
\end{equation}
We refer to the time-dependent unitary transformation constructed in this way as the Schrieffer--Wolff transformation (SWT).

As it stands, the dissipator in Eq.~\eqref{eq:GKSL_eq_for_tilde_rho} is strongly time-dependent and is not of the (time-independent) GKSL form. To remedy this, we expand $\tilde{a}_f(t)$ in a Fourier series,
\begin{equation}
  \tilde{a}_f^I(t) = \sum_{k} L_k\, e^{-i\omega_k t}.
\end{equation}
Substituting this into the dissipator gives
\begin{align}
  &\kappa(n_\text{th}+1)\,\mathcal{D}[\tilde{a}_f^I(t)]\tilde{\rho}^I + \kappa\,n_\text{th}\,\mathcal{D}[(\tilde{a}_f^I)^\dagger(t)]\tilde{\rho}^I \notag\\
  &= \sum_{k,l} e^{-i(\omega_k-\omega_l)t}\Bigl[
  \kappa(n_\text{th}+1)\Bigl(L_k\tilde{\rho}^I L_l^\dagger - \tfrac12\{L_l^\dagger L_k,\tilde{\rho}^I\}\Bigr) 
  + \kappa\,n_\text{th}\Bigl(L_l^\dagger \tilde{\rho}^I L_k - \tfrac12\{L_k L_l^\dagger,\tilde{\rho}^I\}\Bigr)\Bigr].
\end{align}
Assuming that, for $k\neq l$, the inverse detuning $|\omega_k-\omega_l|^{-1}$ is much shorter than the characteristic timescale of the dynamics of interest, we perform the rotating-wave approximation (RWA) and discard the $k\neq l$ terms. 
After returning to the Schr\"odinger picture, we thus arrive at the long-time effective GKSL master equation
\begin{equation}
  \partial_t \tilde{\rho} = -\frac{i}{\hbar}[H_{\mathrm{SWT}}(t),\tilde{\rho}]
    + \sum_{k}\Bigl[\kappa(n_\text{th}+1)\,\mathcal{D}[L_k]\tilde{\rho} + \kappa\,n_\text{th}\,\mathcal{D}[L_k^\dagger]\tilde{\rho}\Bigr],
\end{equation}
where $H_{\mathrm{SWT}}(t) = H_0 + e^{-iH_0t/\hbar}H_{\mathrm{SWT}}^I(t) e^{iH_0t/\hbar}$.

\subsection{Summary of the Schrieffer--Wolff transformation}
We now summarize the outcome of the SWT for the parameter regime in which the engine operates, focusing on the structure of the resulting near-static Hamiltonian $H_{\mathrm{SWT}}$ rather than on the technical details of the expansion. 
Under the resonance condition Eq.~\eqref{eq: resonance condition of CAS}, $H_{\mathrm{SWT}}$ contains the CAS term
\begin{equation}
  H_{\mathrm{CAS}}(t) = \hbar g_{\mathrm{CAS}}\,L_c\,\sigma_+ e^{-i\omega_dt} + \mathrm{h.c.},
  \qquad L_c \equiv (a_1^\dagger)^c a_2^c,
\end{equation}
which coherently swaps $c$ photons between the cavities in conjunction with a qubit flip. Collecting all contributions, the SWT Hamiltonian organizes into
\begin{equation}
  H_{\mathrm{SWT}}(t) = H_0 + H_{\mathrm{LFS}} + H_{\mathrm{NLFS}} + H_{\mathrm{CAS}}(t) + H_f.
\end{equation}
We describe each piece in turn.

\textit{Linear frequency shifts (LFS).---}These are the terms linear in $n_1$, $n_2$, $n_f$, and $\sigma_z$, which renormalize the bare frequencies of each subsystem. They first appear at $\mathcal{O}(\lambda^2)$ and can be written compactly as
\begin{equation}
  H_{\mathrm{LFS}} \equiv \frac{\hbar(\tilde\omega_q-\omega_q)}{2}\sigma_z + \sum_{j=1,2}\hbar(\tilde\omega_j-\omega_j)\,n_j + \hbar(\tilde\omega_f-\omega_f)\,n_f,
\end{equation}
where the tilded frequencies denote the renormalized values; the qubit shift enters at $\mathcal{O}(\lambda^2)$, while the cavity and filter shifts enter at $\mathcal{O}(\lambda^4)$.

\textit{Nonlinear frequency shifts (NLFS).---}These are the higher-than-linear products of $n_1$, $n_2$, $n_f$, and $\sigma_z$. At leading order,
\begin{equation}
  H_{\mathrm{NLFS}} = \sum_{j=1,2}\hbar\chi_j\,n_j\sigma_z + \bigl(\mathcal{O}(\lambda^4)n_1^2 + \mathcal{O}(\lambda^4)n_1 n_2 + \mathcal{O}(\lambda^4)n_2^2\bigr)\hbar\sigma_z + \mathcal{O}(\lambda^6).
\end{equation}
In the regime of interest the total photon number $N \equiv n_1 + n_2$ is conserved and may be treated as a c-number. Choosing the couplings so that $\chi_1 = \chi_2 \equiv \chi$, the $n_j\sigma_z$ term becomes proportional to $N\sigma_z$ and can be absorbed into the LFS through the redefinition
\begin{equation}
  \hat\omega_q \equiv \tilde\omega_q + 2N\chi,
\end{equation}
leaving only the quadratic remainder,
\begin{equation}
  H_{\mathrm{NLFS}} = \bigl(\mathcal{O}(\lambda^4)n_1^2 + \mathcal{O}(\lambda^4)n_1 n_2 + \mathcal{O}(\lambda^4)n_2^2\bigr)\hbar\sigma_z + \mathcal{O}(\lambda^6).
\end{equation}
This cancellation is the condition stated below Eq.~\eqref{eq:effective_GKSL_eq} of the main text, under which the dispersive $a_i^\dagger a_i\sigma_z$ terms drop out.

\textit{CAS coupling.---}The effective CAS coupling $g_{\mathrm{CAS}}$ scales with the perturbative order of the process,
\begin{equation}
  g_{\mathrm{CAS}} \sim \frac{\Omega\,g_1^c g_2^c}{\Delta\omega^{2c}},
\end{equation}
up to corrections of relative order $\mathcal{O}(\lambda^2)$.

\textit{Coupling to the filter.---}The qubit--filter coupling is renormalized to
\begin{equation}
  \tilde g_f \equiv g_f\Bigl(1 - \frac{\Omega^2}{4(\omega_f-\omega_d)(\omega_q-\omega_d)} - \sum_{j=1,2}\frac{g_j^2}{2(\omega_q-\omega_j)(\omega_f-\omega_j)} + \mathcal{O}(\lambda^4)\Bigr),
\end{equation}
and the leading filter term is $H_f = \hbar\tilde g_f(a_f^\dagger\sigma_- + \mathrm{h.c.})$, together with the subleading processes [the $n_j a_f^\dagger\sigma_-$, $n_f\sigma_z$, and $L_c^\dagger a_f\sigma_z$ terms collected in $\mathcal{L}_{\mathrm{int}}$ below].

To make the filter dissipator explicit, we expand the transformed filter operator $\tilde a_f$ using the Baker--Campbell--Hausdorff (BCH) formula~\cite{Wilcox1967},
\begin{align}
  \tilde a_f(t) &= e^{S(t)}a_f e^{-S(t)}
  = \sum_{n=0}^\infty \frac{1}{n!}(\operatorname{ad}_{S(t)})^n(a_f)\notag\\
  &= a_f  - \frac{g_f\Omega}{2(\omega_q-\omega_d)(\omega_f-\omega_d)}\sigma_z e^{-i(\omega_d-\omega_f) t}\notag\\
  &\quad - \sum_{j=1,2}\frac{g_f g_j}{(\omega_q-\omega_j)(\omega_f-\omega_j)}a_j\sigma_z e^{-i(\omega_j-\omega_f) t} + \mathcal{O}(\lambda^3).
\end{align}
Substituting this expansion into the filter dissipators $\mathcal{D}[\tilde a_f(t)]$ and $\mathcal{D}[\tilde a_f^\dagger(t)]$ and applying the RWA, the master equation after the SWT takes the form
\begin{align}
  \label{eq:GKSL_after_SWT}
  \partial_t\tilde\rho &= -\frac{i}{\hbar}[H_{\mathrm{SWT}}(t),\tilde\rho]
  + \kappa(n_\text{th}+1)\,\mathcal{D}[a_f]\tilde\rho + \kappa\,n_\text{th}\,\mathcal{D}[a_f^\dagger]\tilde\rho\notag\\
  &\quad + \gamma_z(2n_\text{th}+1)\bigl(\sigma_z\tilde\rho\sigma_z - \tilde\rho\bigr)\notag\\
  &\quad + \sum_{j=1,2}\Bigl[\gamma_{\mathrm{loss},j}(n_\text{th}+1)\Bigl(a_j\sigma_z\tilde\rho\sigma_z a_j^\dagger - \tfrac12\{a_j^\dagger a_j,\tilde\rho\}\Bigr)\notag\\
  &\hspace{4.2em} + \gamma_{\mathrm{loss},j}\,n_\text{th}\Bigl(a_j^\dagger\sigma_z\tilde\rho\sigma_z a_j - \tfrac12\{a_j a_j^\dagger,\tilde\rho\}\Bigr)\Bigr] + \mathcal{O}(\kappa g_f^2\lambda^4),
\end{align}
where
\begin{equation}
  \gamma_z = \frac{\kappa g_f^2\Omega^2}{4(\omega_q-\omega_d)^2(\omega_f-\omega_d)^2} + \mathcal{O}(\kappa g_f^2\lambda^4),
  \qquad
  \gamma_{\mathrm{loss},j} = \frac{\kappa g_f^2 g_j^2}{(\omega_q-\omega_j)^2(\omega_f-\omega_j)^2} + \mathcal{O}(\kappa g_f^2\lambda^4).
\end{equation}
The $\gamma_z$ term is a pure-dephasing channel for the qubit and the $\gamma_{\text{loss}}$ term is a photon-loss channel via Purcell filter.
Equation~\eqref{eq:GKSL_after_SWT} is the starting point for the Nakajima--Zwanzig reduction below. 

\subsection{Nakajima--Zwanzig projection formalism}
We first derive a generalized master equation using the Nakajima--Zwanzig projection-operator formalism~\cite{Nakajima1958,Zwanzig1960}, and then reduce it to an effective GKSL equation for the working medium (cavity 1 + cavity 2) under the assumption that the environment (qubit + filter) correlation time is much shorter than the characteristic timescale of the engine dynamics. Throughout this subsection, we label operators acting only on the working medium by the subscript $12$ and those acting only on the environment by the subscript $qf$.

We write the dynamics of the total state $\tilde\rho$ after the SWT~[Eq.~\eqref{eq:GKSL_after_SWT}] as
\begin{equation}
  \partial_t\tilde\rho = \mathcal{L}_{12}\,\tilde\rho + \mathcal{L}_{\mathrm{int}}(t)\,\tilde\rho + \mathcal{L}_{qf}\,\tilde\rho,
\end{equation}
where $\mathcal{L}_{12}$ and $\mathcal{L}_{qf}$ are GKSL generators acting on the working medium and the environment, respectively, and $\mathcal{L}_{\mathrm{int}}(t)$ describes their interaction; only $\mathcal{L}_{\mathrm{int}}(t)$ carries an explicit time dependence, through the drive. We define the reduced state of the working medium by the partial trace over the environment,
\begin{equation}
  \rho_{12} \equiv \Tr_{qf}[\tilde\rho],
\end{equation}
and denote by $\rho_{qf}^{\mathrm{ss}}$ the stationary environment state, $\mathcal{L}_{qf}\,\rho_{qf}^{\mathrm{ss}} = 0$. Using $\rho_{qf}^{\mathrm{ss}}$, we introduce the projection superoperator $\mathcal{P}$ and its complement $\mathcal{Q}$,
\begin{equation}
  \mathcal{P}X \equiv \Tr_{qf}[X]\otimes\rho_{qf}^{\mathrm{ss}},
  \qquad
  \mathcal{Q} \equiv \mathbf{1} - \mathcal{P},
\end{equation}
for an arbitrary operator $X$ on the total space, so that $\mathcal{P}+\mathcal{Q}=\mathbf{1}$ and $\mathcal{P}\tilde\rho(t) = \rho_{12}(t)\otimes\rho_{qf}^{\mathrm{ss}}$.

Passing to the interaction picture with respect to $\mathcal{L}_{12}+\mathcal{L}_{qf}$,
\begin{equation}
  \tilde\rho^{\,\mathcal{I}}(t) = e^{-(\mathcal{L}_{12}+\mathcal{L}_{qf})t}\,\tilde\rho(t),
\end{equation}
we have $\mathcal{P}\tilde\rho^{\,\mathcal{I}}(t) \equiv \rho_{12}^{\mathcal{I}}(t)\otimes\rho_{qf}^{\mathrm{ss}} = e^{-\mathcal{L}_{12}t}\rho_{12}(t)\otimes\rho_{qf}^{\mathrm{ss}}$, so it suffices to determine $\tilde\rho^{\mathcal{I}}(t)$, which obeys
\begin{equation}
  \partial_t\tilde\rho^{\mathcal{I}} = \mathcal{L}_{\mathrm{int}}^{\mathcal{I}}(t)\,\tilde\rho^{\mathcal{I}}(t),
  \qquad
  \mathcal{L}_{\mathrm{int}}^{\mathcal{I}}(t) = e^{-(\mathcal{L}_{12}+\mathcal{L}_{qf})t}\,\mathcal{L}_{\mathrm{int}}(t)\,e^{(\mathcal{L}_{12}+\mathcal{L}_{qf})t}.
\end{equation}
Since $\mathcal{P}$ is time independent, projecting with $\mathcal{P}$ and tracing over the environment gives
\begin{equation}
  \label{eq:rho12_with_Qrho}
  \partial_t\rho_{12}^{\mathcal{I}} = \Tr_{qf}\Bigl[\mathcal{L}_{\mathrm{int}}^{\mathcal{I}}(t)\bigl(\rho_{12}^{\mathcal{I}}(t)\otimes\rho_{qf}^{\mathrm{ss}}\bigr) + \mathcal{L}_{\mathrm{int}}^{\mathcal{I}}(t)\,\mathcal{Q}\tilde\rho^{\mathcal{I}}(t)\Bigr],
\end{equation}
which is not closed in $\rho_{12}^{\mathcal{I}}$ because of the $\mathcal{Q}\tilde\rho^{\mathcal{I}}$ term. Projecting instead with $\mathcal{Q}$ and integrating,
\begin{equation}
  \mathcal{Q}\tilde\rho^{\mathcal{I}}(t) = \mathcal{Q}\tilde\rho^{\mathcal{I}}(0) + \int_0^t\! dt_1\,\Bigl(\mathcal{Q}\mathcal{L}_{\mathrm{int}}^{\mathcal{I}}(t_1)\mathcal{P}\tilde\rho^{\mathcal{I}}(t_1) + \mathcal{Q}\mathcal{L}_{\mathrm{int}}^{\mathcal{I}}(t_1)\mathcal{Q}\tilde\rho^{\mathcal{I}}(t_1)\Bigr).
\end{equation}
Iterating the substitution of $\mathcal{Q}\tilde\rho^{\mathcal{I}}$ on the right-hand side, we obtain
\begin{equation}
  \label{eq:Qrho_integral_rep}
  \mathcal{Q}\tilde{\rho}^{\mathcal{I}}(t)=\Lambda(t;0)\mathcal{Q}\tilde{\rho}^{\mathcal{I}}(0)+\int_0^t\,ds \Lambda(t;s)\mathcal{Q}\mathcal{L}_{\text{int}}^{\mathcal{I}}(s)\mathcal{P}\tilde{\rho}^{\mathcal{I}}(s),
\end{equation}
Here, we define the propagator
\begin{equation}
  \label{eq:Lambda}
  \Lambda(t;s) = \mathcal{T}_t\Bigl[\exp\Bigl(\int_s^t\! du\,\mathcal{Q}\mathcal{L}_{\mathrm{int}}^{\mathcal{I}}(u)\Bigr)\Bigr],
\end{equation}
where $\mathcal{T}_t$ is the time-ordering operator, arranging superoperators chronologically from right to left. 
Inserting Eq.~\eqref{eq:Qrho_integral_rep} into Eq.~\eqref{eq:rho12_with_Qrho}, we obtain the generalized master equation in the interaction picture,
\begin{align}
  \label{eq:generalized_master_eq_I}
  \partial_t\rho_{12}^{\mathcal{I}}(t)
  &= \Tr_{qf}\Bigl[\mathcal{L}_{\mathrm{int}}^{\mathcal{I}}(t)\bigl(\rho_{12}^{\mathcal{I}}(t)\otimes\rho_{qf}^{\mathrm{ss}}\bigr)\Bigr]
   + \Tr_{qf}\Bigl[\mathcal{L}_{\mathrm{int}}^{\mathcal{I}}(t)\Lambda(t;0)\mathcal{Q}\tilde\rho^{\mathcal{I}}(0)\Bigr]\notag\\
  &\quad + \int_0^t\! ds\,\Tr_{qf}\Bigl[\mathcal{L}_{\mathrm{int}}^{\mathcal{I}}(t)\Lambda(t;s)\mathcal{Q}\mathcal{L}_{\mathrm{int}}^{\mathcal{I}}(s)\bigl(\rho_{12}^{\mathcal{I}}(s)\otimes\rho_{qf}^{\mathrm{ss}}\bigr)\Bigr].
\end{align}
Acting with $e^{\mathcal{L}_{12}t}$ to return to the Schr\"odinger picture,
\begin{align}
  \label{eq:generalized_master_eq}
  \partial_t\rho_{12}(t)
  &= \mathcal{L}_{12}\rho_{12}(t) + \Tr_{qf}\Bigl[\mathcal{L}_{\mathrm{int}}(t)\bigl(\rho_{12}(t)\otimes\rho_{qf}^{\mathrm{ss}}\bigr)\Bigr]+ e^{\mathcal{L}_{12}t}\Tr_{qf}\Bigl[\mathcal{L}_{\mathrm{int}}^{\mathcal{I}}(t)\Lambda(t;0)\mathcal{Q}\tilde\rho(0)\Bigr]\notag\\
  &\quad + e^{\mathcal{L}_{12}t}\int_0^t\! ds\,\Tr_{qf}\Bigl[\mathcal{L}_{\mathrm{int}}^{\mathcal{I}}(t)\Lambda(t;s)\mathcal{Q}\mathcal{L}_{\mathrm{int}}^{\mathcal{I}}(s)\bigl(e^{-\mathcal{L}_{12}s}\rho_{12}(s)\otimes\rho_{qf}^{\mathrm{ss}}\bigr)\Bigr].
\end{align}
We refer to Eq.~\eqref{eq:generalized_master_eq} as the generalized master equation; no approximation has been made up to this point.

After the SWT, the generators of Eq.~\eqref{eq:GKSL_after_SWT} split into a working-medium part $\mathcal{L}_{12}$, an environment part $\mathcal{L}_{qf}$, and their interaction $\mathcal{L}_{\mathrm{int}}(t)$. Using the SWT Hamiltonian summarized in the last subsection, at leading order
\begin{align}
  \mathcal{L}_{12}\,\rho &= -\frac{i}{\hbar}\Bigl[\sum_{j=1,2}\hbar\tilde\omega_j n_j,\,\rho\Bigr] + \mathcal{O}(\lambda^4),\\
  \mathcal{L}_{qf}\,\rho &= -\frac{i}{\hbar}\Bigl[\frac{\hbar\hat\omega_q}{2}\sigma_z + \hbar\tilde\omega_f a_f^\dagger a_f + \hbar\tilde g_f\bigl(a_f^\dagger\sigma_- + a_f\sigma_+\bigr),\,\rho\Bigr]\notag\\
  &\quad + \kappa(n_\text{th}+1)\,\mathcal{D}[a_f]\rho + \kappa\,n_\text{th}\,\mathcal{D}[a_f^\dagger]\rho + \mathcal{O}(g_f\lambda^2),
\end{align}
while $\mathcal{L}_{\mathrm{int}}(t)$ collects the CAS term together with the nonlinear dispersive shift, filter-coupling, dephasing, and photon-loss terms,
\begin{align}
  \label{eq:Lint}
  \mathcal{L}_{\mathrm{int}}(t)\,\rho
  &= -\frac{i}{\hbar}\bigl[\hbar g_{\mathrm{CAS}}\,L_c^\dagger\,\sigma_- e^{i\omega_d t} + \mathrm{h.c.},\,\rho\bigr] + \mathcal{O}(\lambda^{2c+3})\notag\\
  &\quad -\frac{i}{\hbar}\bigl[\bigl(\mathcal{O}(\lambda^4)n_1^2 + \mathcal{O}(\lambda^4)n_2^2 + \mathcal{O}(\lambda^4)n_1 n_2\bigr)\hbar\sigma_z,\,\rho\bigr] + \mathcal{O}(\lambda^6)\notag\\
  &\quad -\frac{i}{\hbar}\Bigl[-\sum_{j=1,2}\mathcal{O}(g_f\lambda^2)\,\hbar n_j a_f^\dagger\sigma_- e^{-i(\omega_q-\omega_f)t} + \mathrm{h.c.},\,\rho\Bigr] + \mathcal{O}(g_f\lambda^4)\notag\\
  &\quad -\frac{i}{\hbar}\bigl[\mathcal{O}(g_f\lambda^{2c+1})\,L_c^\dagger a_f\sigma_z e^{-i(\omega_f-\omega_d-c(\omega_2-\omega_1))t} + \mathrm{h.c.},\,\rho\bigr] + \mathcal{O}(g_f\lambda^{2c+3})\notag\\
  &\quad +\gamma_z (2n_\text{th}+1) (\sigma_z \rho \sigma_z -\rho)\notag \\
  &\quad + \sum_{j=1,2}\Bigl[\gamma_{\mathrm{loss},j}(n_\text{th}+1)\Bigl(a_j\sigma_z\rho\sigma_z a_j^\dagger - \tfrac12\{a_j^\dagger a_j,\rho\}\Bigr)\notag\\
  &\hspace{4.2em} + \gamma_{\mathrm{loss},j}\,n_\text{th}\Bigl(a_j^\dagger\sigma_z\rho\sigma_z a_j - \tfrac12\{a_j a_j^\dagger,\rho\}\Bigr)\Bigr] + \mathcal{O}(\kappa g_f^2\lambda^4),
\end{align}
where $\gamma_{\mathrm{loss},j}$ are the residual photon-loss rates. Although the last two lines of $\mathcal{L}_{\mathrm{int}}$ act only on the working medium, we include them in the interaction generator for convenience. For $g_f\ll\kappa$ the stationary environment state is approximately incoherent,
\begin{equation}
  \rho_{qf}^{\mathrm{ss}} \underset{\sim}{\propto} e^{-\beta\hbar\hat\omega_q\sigma_z/2}\otimes e^{-\beta\hbar\tilde\omega_f n_f},
\end{equation}
which we use throughout.

We now examine the three contributions to Eq.~\eqref{eq:generalized_master_eq}.

\textit{(i) Time-local term.---}The first line, $\mathcal{L}_{12}\rho_{12}(t) + \Tr_{qf}[\mathcal{L}_{\mathrm{int}}(t)(\rho_{12}(t)\otimes\rho_{qf}^{\mathrm{ss}})]$, renormalizes the Hamiltonian and the dissipator of the working medium by replacing the environment operators in $\mathcal{L}_{\mathrm{int}}$ by their stationary expectation values. Since $\rho_{qf}^{\mathrm{ss}}$ has no coherences, the off-diagonal contributions vanish, leaving
\begin{align}
  &\mathcal{L}_{12}\rho_{12} + \Tr_{qf}\Bigl[\mathcal{L}_{\mathrm{int}}(t)\bigl(\rho_{12}\otimes\rho_{qf}^{\mathrm{ss}}\bigr)\Bigr]\notag\\
  &= -i\Bigl[\sum_{j=1,2}\tilde\omega_j n_j,\,\rho_{12}\Bigr] - i\Bigl[\bigl(\mathcal{O}(\lambda^4)n_1^2 + \mathcal{O}(\lambda^4)n_2^2 + \mathcal{O}(\lambda^4)n_1 n_2\bigr)\langle\sigma_z\rangle_{\mathrm{ss}},\,\rho_{12}\Bigr] + \mathcal{O}(\lambda^6)\notag\\
  &\quad + \sum_{j=1,2}\Bigl[\gamma_{\mathrm{loss},j}(n_\text{th}+1)\,\mathcal{D}[a_j]\rho_{12} + \gamma_{\mathrm{loss},j}\,n_\text{th}\,\mathcal{D}[a_j^\dagger]\rho_{12}\Bigr] + \mathcal{O}(\kappa g_f^2\lambda^4),
\end{align}
where $\langle\sigma_z\rangle_{\mathrm{ss}} \equiv \Tr(\sigma_z\rho_{qf}^{\mathrm{ss}})$ and we used $\sigma_z^2=\mathbf{1}$.

\textit{(ii) Initial-slip term.---}The second line vanishes for a factorized initial state, $\tilde\rho(0) = \rho_{12}(0)\otimes\rho_{qf}^{\mathrm{ss}}$. Otherwise---if the initial qubit state deviates from $\rho_{qf}^{\mathrm{ss}}$, or initial medium--environment correlations are present---it is nonzero, but transient, and decays after a sufficiently long time.

\textit{(iii) Memory term.---}The third line describes how deviations from the environment steady state, generated at earlier times $s<t$ by $\mathcal{L}_{\mathrm{int}}^{I}(s)$, feed back into the working-medium dynamics at time $t$. In the next subsection we approximate this time-nonlocal term by a time-local one, reducing the generalized master equation to a Markovian GKSL form.

\subsection{Adiabatic elimination of the Purcell filter}
To reduce the generalized master equation [Eq.~\eqref{eq:generalized_master_eq}] to a time-local form, the central objects are the environment correlation functions. We focus on
\begin{equation}
  \Phi_{\pm\mp}(t) \equiv \Tr_{qf}\bigl[\sigma_\pm\, e^{\mathcal{Q}\mathcal{L}_{qf}t}\,\mathcal{Q}\sigma_\mp\,\rho_{qf}^{\mathrm{ss}}\bigr],
  \qquad
  \Phi_{zz}(t) \equiv \Tr_{qf}\bigl[\delta\sigma_z\, e^{\mathcal{Q}\mathcal{L}_{qf}t}\,\mathcal{Q}\delta\sigma_z\,\rho_{qf}^{\mathrm{ss}}\bigr],
\end{equation}
with the fluctuation $\delta\sigma_z \equiv \sigma_z - \langle\sigma_z\rangle_{\mathrm{ss}}$. Neglecting the $n_f\sigma_z$ term for $g_f$ sufficiently small, the environment generator reads, at leading order,
\begin{align}
  \mathcal{L}_{qf}\,\rho &= -\frac{i}{\hbar}\Bigl[\frac{\hbar\hat\omega_q}{2}\sigma_z + \hbar\tilde\omega_f a_f^\dagger a_f + \hbar\tilde g_f\bigl(a_f^\dagger\sigma_- + a_f\sigma_+\bigr),\,\rho\Bigr]\notag\\
  &\quad + \kappa(n_\text{th}+1)\,\mathcal{D}[a_f]\rho + \kappa\,n_\text{th}\,\mathcal{D}[a_f^\dagger]\rho + \mathcal{O}(g_f\lambda^2).
\end{align}

It is convenient to pass to the Heisenberg picture. Equipping the operator space with the Hilbert--Schmidt inner product $\langle X,Y\rangle \equiv \Tr(X^\dagger Y)$, the adjoint generator $\mathcal{L}_{qf}^\dagger$ is defined by $\langle X,\mathcal{L}_{qf}Y\rangle = \langle Y,\mathcal{L}_{qf}^\dagger X\rangle^*$ for all $X,Y$, and reads
\begin{align}
  \mathcal{L}_{qf}^\dagger\,X &= \frac{i}{\hbar}\Bigl[\frac{\hbar\hat\omega_q}{2}\sigma_z + \hbar\tilde\omega_f a_f^\dagger a_f + \hbar\tilde g_f\bigl(a_f^\dagger\sigma_- + a_f\sigma_+\bigr),\,X\Bigr]\notag\\
  &\quad + \kappa(n_\text{th}+1)\Bigl(a_f^\dagger X a_f - \tfrac12\{a_f^\dagger a_f,X\}\Bigr) + \kappa\,n_\text{th}\Bigl(a_f X a_f^\dagger - \tfrac12\{a_f a_f^\dagger,X\}\Bigr).
\end{align}
Writing $X(t) \equiv e^{\mathcal{L}_{qf}^\dagger t}X$ for the Heisenberg-picture operators, the leading-order equations of motion are
\begin{equation}
\begin{cases}
  \partial_t a_f(t) = -\bigl(i\tilde\omega_f + \tfrac{\kappa}{2}\bigr)a_f(t) - i\tilde g_f\,\sigma_-(t),\\[2pt]
  \partial_t[a_f\sigma_z](t) = -\bigl(i\tilde\omega_f + \tfrac{\kappa}{2}\bigr)[a_f\sigma_z](t) + i\tilde g_f\bigl([(2n_f+1)\sigma_-](t) - 2a_f^2\sigma_+(t)\bigr),\\[2pt]
  \partial_t[a_f\sigma_+](t) = -\bigl(-i(\hat\omega_q-\tilde\omega_f) + \tfrac{\kappa}{2}\bigr)[a_f\sigma_+](t) - i\tilde g_f\bigl([n_f\sigma_z](t) + \tfrac{\sigma_z(t)+1}{2}\bigr),\\[2pt]
  \partial_t\sigma_-(t) = -i\hat\omega_q\,\sigma_-(t) + i\tilde g_f\,[a_f\sigma_z](t),\\[2pt]
  \partial_t\sigma_z(t) = 2i\tilde g_f\bigl([a_f^\dagger\sigma_-](t) - [a_f\sigma_+](t)\bigr).
\end{cases}
\label{eq:Heisenberg_eq_afs}
\end{equation}
These do not close on themselves. Integrating the equation for $a_f$ and resumming,
\begin{equation}
  a_f(t) = e^{-(i\tilde\omega_f + \frac{\kappa}{2})t}a_f - i\tilde g_f\int_0^t\! ds\, e^{-(i\tilde\omega_f + \frac{\kappa}{2})(t-s)}\sigma_-(s),
\end{equation}
and similarly for $[a_f\sigma_z](t)$ and $[a_f\sigma_+](t)$. For $g_f\ll\kappa$, the interaction is negligible on the timescale $t-s\lesssim\kappa^{-1}$, so the slow operators may be taken outside the integral. Together with $[n_f\sigma_-](t)\simeq n_\text{th}\sigma_-(t)$, $[a_f^2\sigma_+](t)\approx 0$, and $[n_f\sigma_z](t)\simeq n_\text{th}\sigma_z(t)$ for $t\gg\kappa^{-1}$, this gives
\begin{align}
  [a_f\sigma_z](t) &\approx e^{-(i\tilde\omega_f + \frac{\kappa}{2})t}a_f\sigma_z + \frac{i\tilde g_f}{\frac{\kappa}{2} - i(\hat\omega_q-\tilde\omega_f)}(2n_\text{th}+1)\sigma_-(t),\\
  [a_f\sigma_+](t) &\approx e^{-(-i(\hat\omega_q-\tilde\omega_f) + \frac{\kappa}{2})t}a_f\sigma_+ - \frac{i\tilde g_f}{\frac{\kappa}{2} - i(\hat\omega_q-\tilde\omega_f)}\bigl(n_\text{th}+\tfrac12\bigr)\delta\sigma_z(t),
\end{align}
where we used $\langle\sigma_z\rangle_{\mathrm{ss}} = \Tr(\sigma_z\rho_{qf}^{\mathrm{ss}}) = -(2n_\text{th}+1)^{-1}$. Substituting these back into Eq.~\eqref{eq:Heisenberg_eq_afs} and introducing
\begin{equation}
  \alpha_P = \gamma_P + i\omega_P \equiv \frac{\tilde g_f^2}{\frac{\kappa}{2} - i(\hat\omega_q-\tilde\omega_f)},
\end{equation}
the qubit operators relax as
\begin{align}
  \label{eq:approximated_sigma_minus}
  \sigma_-(t) &\approx e^{(-i\hat\omega_q - (2n_\text{th}+1)\alpha_P)t}\Bigl[\sigma_- + i\tfrac{\alpha_P}{\tilde g_f}a_f\sigma_z\bigl(1 - e^{-(-i(\hat\omega_q-\tilde\omega_f) + \frac{\kappa}{2})t}\bigr)\Bigr],\\
  \label{eq:approximated_sigma_z}
  \sigma_z(t) &\approx \langle\sigma_z\rangle_{\mathrm{ss}} + e^{-(2n_\text{th}+1)2\gamma_P t}\Bigl[\delta\sigma_z - 2i\tfrac{\alpha_P}{\tilde g_f}a_f\sigma_+\bigl(1 - e^{-(-i(\hat\omega_q-\tilde\omega_f) + \frac{\kappa}{2})t}\bigr)\Bigr].
\end{align}
The correlation functions then reduce to Heisenberg-picture expectation values, $\Phi_{\pm\mp}(t) = \Tr[\sigma_\pm(t)\sigma_\mp\rho_{qf}^{\mathrm{ss}}]$ and $\Phi_{zz}(t) = \Tr[\delta\sigma_z(t)\delta\sigma_z\rho_{qf}^{\mathrm{ss}}] $, yielding
\begin{align}
  \label{eq:bath_correlation_functions}
  \Phi_{-+}(t) &\approx e^{(-i\hat\omega_q - (2n_\text{th}+1)\alpha_P)t}\,\langle g|\rho_{qf}^{\mathrm{ss}}|g\rangle + \mathcal{O}\!\bigl(\tfrac{g_f}{\kappa}\bigr),\\
  \Phi_{+-}(t) &\approx e^{(i\hat\omega_q - (2n_\text{th}+1)\alpha_P^*)t}\,\langle e|\rho_{qf}^{\mathrm{ss}}|e\rangle + \mathcal{O}\!\bigl(\tfrac{g_f}{\kappa}\bigr),\\
  \Phi_{zz}(t) &\approx e^{-(2n_\text{th}+1)2\gamma_P t}\bigl(1 - \langle\sigma_z\rangle_{\mathrm{ss}}^2\bigr) + \mathcal{O}\!\bigl(\tfrac{g_f}{\kappa}\bigr).
\end{align}
We refer to this sequence of approximations collectively as the adiabatic elimination of the Purcell filter~\cite{Sete2015}.

\subsection{Born--Markov approximation and the effective GKSL equation}
We now return to the memory term (iii) of Eq.~\eqref{eq:generalized_master_eq}. Expanding the propagator $\Lambda(t;s)$ of Eq.~\eqref{eq:Lambda} and introducing $\widetilde{\mathcal{L}}_{\mathrm{int}}(t) \equiv e^{-\mathcal{L}_{12}t}\mathcal{L}_{\mathrm{int}}(t)e^{\mathcal{L}_{12}t}$, the leading CAS contribution reads
\begin{equation}
  \widetilde{\mathcal{L}}_{\mathrm{int}}(t)\,\rho = -\frac{i}{\hbar}\bigl[\hbar g_{\mathrm{CAS}}\,L_c^\dagger\,\sigma_- e^{i(\hat{\omega}_q+\delta\tilde\omega) t} + \mathrm{h.c.},\,\rho\bigr] + \cdots,
  \qquad
  \delta\tilde\omega \equiv \omega_d + c(\tilde\omega_2-\tilde\omega_1) - \hat\omega_q,
\end{equation}
where the ellipsis collects the nonlinear dispersive shift, filter-coupling, dephasing, and photon-loss terms. These terms are negligible for the following reasons. The filter-coupling terms relax at rate $\mathcal{O}(\kappa)$ and are therefore suppressed at long times for $\gamma_P\ll\kappa$. The dephasing terms vanish because $\rho^{\mathrm{ss}}_{qf}$ approximately commutes with $\sigma_z$. The photon-loss terms lie in $\ker\mathcal{Q}$ when acting on $\rho_{12}\otimes\rho_{qf}^{\mathrm{ss}}$. 
The nonlinear dispersive shift produces dephasing of the working medium, but this contribution is also negligible in the operating regime of the engine, where the coherences of the working medium are near zero.
The CAS term therefore dominates.
Keeping it, $\widetilde{\mathcal{L}}_{\mathrm{int}}(t)$ factorizes as
\begin{equation}
  \widetilde{\mathcal{L}}_{\mathrm{int}}(t) = \sum_{k}\mathcal{A}_k\otimes\mathcal{B}_k^t,
\end{equation}
with working-medium superoperators $\mathcal{A}_{-,L}\rho = L_c^\dagger\rho$, $\mathcal{A}_{-,R}\rho = \rho L_c^\dagger$, $\mathcal{A}_{+,L}\rho = L_c\rho$, $\mathcal{A}_{+,R}\rho = \rho L_c$, and environment superoperators $\mathcal{B}_{-,L}^t\rho_{qf} = -i g_{\mathrm{CAS}}e^{i(\hat\omega_q+\delta\tilde\omega)t}\sigma_-\rho_{qf}$, $\mathcal{B}_{-,R}^t\rho_{qf} = i g_{\mathrm{CAS}}e^{i(\hat\omega_q+\delta\tilde\omega)t}\rho_{qf}\sigma_-$ (and Hermitian conjugates for $k=+$). The memory term becomes a nested series whose kernels are the correlation functions $\Phi_{\pm\mp}$. Using Eqs.~\eqref{eq:approximated_sigma_minus} and~\eqref{eq:bath_correlation_functions}, the $n$-th term scales as $(g_{\mathrm{CAS}}/\gamma_P)^n$, so under the Born condition
\begin{equation}
  \label{eq:Born_condition}
  \|\mathcal{A}\| \approx \sqrt{\overline{n_1}^c\,\overline{n_2}^c} \;\ll\; (g_{\mathrm{CAS}}/\gamma_P)^{-1}
\end{equation}
(with $\overline{n_j}$ a typical photon number) we may keep only the lowest order $n=0$,
\begin{align}
  \label{eq:Born_approximation}
  (\mathrm{iii}) &\approx g_{\mathrm{CAS}}^2\int_0^t\! ds\,\Bigl[
    e^{-i(\hat\omega_q+\delta\tilde\omega)(t-s)}\Phi_{+-}(t-s)\bigl(L_c^\dagger\rho_{12}(s)L_c - L_c L_c^\dagger\rho_{12}(s)\bigr)\notag\\
  &\hspace{5.4em} + e^{i(\hat\omega_q+\delta\tilde\omega)(t-s)}\Phi_{+-}(t-s)^*\bigl(L_c^\dagger\rho_{12}(s)L_c - \rho_{12}(s)L_c L_c^\dagger\bigr)\notag\\
  &\hspace{5.4em} + e^{-i(\hat\omega_q+\delta\tilde\omega)(t-s)}\Phi_{-+}(t-s)^*\bigl(L_c\rho_{12}(s)L_c^\dagger - L_c^\dagger L_c\rho_{12}(s)\bigr)\notag\\
  &\hspace{5.4em} + e^{i(\hat\omega_q+\delta\tilde\omega)(t-s)}\Phi_{-+}(t-s)\bigl(L_c\rho_{12}(s)L_c^\dagger - \rho_{12}(s)L_c^\dagger L_c\bigr)\Bigr].
\end{align}
We refer to this as the Born approximation.

Since $\Phi_{\pm\mp}(t)$ decays at rate $(2n_\text{th}+1)\gamma_P$, under the separation of timescales $\tau_s^{-1}\ll(2n_\text{th}+1)\gamma_P$ we may replace $\rho_{12}(s)\to\rho_{12}(t)$, and for $t\gg[(2n_\text{th}+1)\gamma_P]^{-1}$ extend the lower limit to $-\infty$,
\begin{equation}
  \label{eq:Phi_pm_approx}
  \int_0^t\! ds\, e^{\mp i(\hat\omega_q+\delta\tilde\omega)(t-s)}\Phi_{\pm\mp}(t-s)
  \approx \Bigl(\frac{\Gamma_{q}}{2}\pm iS_{q}\Bigr)\times
  \begin{cases}
    \langle e|\rho_{qf}^{\mathrm{ss}}|e\rangle & (\Phi_{+-}),\\
    \langle g|\rho_{qf}^{\mathrm{ss}}|g\rangle & (\Phi_{-+}).
  \end{cases}
\end{equation}
At resonance $\tilde\omega_f = \hat\omega_q$ and $\delta\tilde\omega = 0$, this gives
\begin{equation}
  \label{eq:Gamma_q}
  \Gamma_{q} = \frac{2}{(2n_\text{th}+1)\gamma_P} = \frac{\kappa}{(2n_\text{th}+1)\tilde g_f^2},\qquad S_{q} = 0,
\end{equation}
on which we focus hereafter. We also drop the initial-slip term (ii), which is transient on the same timescale (or vanishes exactly for a factorized initial state). We refer to these steps collectively as the Markov approximation, and to the combination with the Born approximation as the Born--Markov approximation.

Under these approximations, Eq.~\eqref{eq:generalized_master_eq} reduces to the effective GKSL equation for the working medium,
\begin{equation}
  \label{eq:effective_GKSL_supp}
  \partial_t\rho_{12} = -\frac{i}{\hbar}\bigl[H_{\mathrm{eff}},\rho_{12}\bigr]
    + \gamma_{2\to1}\,\mathcal{D}[L_c]\rho_{12} + \gamma_{1\to2}\,\mathcal{D}[L_c^\dagger]\rho_{12}
    + \sum_{j=1,2}\gamma_{\mathrm{loss},j}\Bigl[(n_\text{th}+1)\mathcal{D}[a_j]\rho_{12} + n_\text{th}\,\mathcal{D}[a_j^\dagger]\rho_{12}\Bigr],
\end{equation}
with $H_{\mathrm{eff}} = \sum_{j=1,2}\hbar\tilde\omega_j n_j + (\mathcal{O}(\lambda^4)n_1^2 + \mathcal{O}(\lambda^4)n_1 n_2 + \mathcal{O}(\lambda^4)n_2^2)\hbar\langle\sigma_z\rangle_{\mathrm{ss}} + \mathcal{O}(\lambda^6)$ and CAS rates
\begin{equation}
  \gamma_{2\to1} \approx \Gamma_{q}\,g_{\mathrm{CAS}}^2\,\langle g|\rho_{qf}^{\mathrm{ss}}|g\rangle,
  \qquad
  \gamma_{1\to2} \approx \Gamma_{q}\,g_{\mathrm{CAS}}^2\,\langle e|\rho_{qf}^{\mathrm{ss}}|e\rangle.
\end{equation}
Equation~\eqref{eq:effective_GKSL_supp} is precisely Eq.~\eqref{eq:effective_GKSL_eq} of the main text; the detailed-balance ratio $\gamma_{2\to1}/\gamma_{1\to2} \approx \langle g|\rho_{qf}^{\mathrm{ss}}|g\rangle/\langle e|\rho_{qf}^{\mathrm{ss}}|e\rangle = e^{\beta\hbar\hat\omega_q}\approx e^{\beta \hbar \omega_q}$ fixes the effective inverse temperature $\beta_{\mathrm{eff}}$ used there.

\subsection{Validity regime of the approximation}
We summarize the assumptions underlying the derivation. Since the engine operates near thermal equilibrium, where the lower cavity is predominantly occupied, we estimate the typical photon numbers as $\overline{n_1} = \mathcal{O}(N)$ and $\overline{n_2} = \mathcal{O}(1)$, so that $\overline{n_1}^c\,\overline{n_2}^c = \mathcal{O}(N^c)$.

\textit{(i) Weak coupling and weak driving.---}
\begin{equation}
  \frac{N^{\frac12}g_j}{\Delta\omega},\ \frac{\Omega}{\Delta\omega} \;\ll\; 1.
\end{equation}

\textit{(ii) Adiabatic elimination of the Purcell filter.---}$g_f\ll\kappa$.

\textit{(iii) Born approximation.---}$N^c\,g_{\mathrm{CAS}}^2 \ll \gamma_P^2$.

\textit{(iv) Markov approximation.---}$\tau_s^{-1}\ll\gamma_P$ and $\gamma_P^{-1}\ll t$.

To make the CAS process dominate the working-medium dynamics, we further impose

\textit{(v) CAS-dominance condition.---}$N\gamma_{\mathrm{loss},1,2} \ll N^{c}\Gamma_q\,g_{\mathrm{CAS}}^2$.

Under (v) the characteristic timescale is set by the CAS transitions, $\tau_s^{-1}\approx N^c\Gamma_q g_{\mathrm{CAS}}^2$, so combining (v) with (iii) and using $\Gamma_q\approx\gamma_P^{-1}$ yields $N\gamma_P\gamma_{\mathrm{loss},1,2}\ll N^{c}g_{\mathrm{CAS}}^2\ll\gamma_P^2$.
With the estimates $\gamma_{\mathrm{loss},1,2}\approx\kappa g_f^2 g_{1,2}^2/\Delta\omega^4$, $\gamma_P\approx g_f^2/\kappa$, and $g_{\mathrm{CAS}}\approx\Omega g_1^c g_2^c/\Delta\omega^{2c}$, together with (ii), this yields the hierarchical window,
\begin{equation}
  \frac{N^{\frac12}g_f^2g_{1,2}}{\Delta\omega^2} \;\ll\; \Omega\Bigl(\frac{N^{\frac12} g_1 g_2}{\Delta\omega^2}\Bigr)^{c} \;\ll\; \frac{g_f^2}{\kappa}, \quad g_f\ll \kappa.
\end{equation}
whose first two inequalities are Eq.~\eqref{eq:approximation_condition} of the main text.

We have numerically confirmed that the parameter dependence of Eq.~\eqref{eq:Gamma_q} remains approximately valid even outside the regime $g_f\ll\kappa$. 
In any case, this approximation is not actually used in our simulations: rather than relying on the analytic estimate [Eq.~\eqref{eq:Gamma_q}], we evaluate $\Gamma_q$ by computing the correlation functions directly from $\mathcal{L}_{qf}$.
It therefore suffices to restrict the parameter regime by the hierarchical window [Eq.~\eqref{eq:approximation_condition}] alone.

\section{Heat and work from the underlying unitary dynamics}
\label{sec:heat_and_work_unitary}
In the main text, heat and work for the working medium were defined through the second law and the first law applied to the effective dynamics [Eq.~\eqref{eq:coherent_GKSL_eq}]. Here we show that the same definitions follow from the underlying unitary dynamics and Markov approximation, providing an independent justification of Eqs.~\eqref{eq:heat_current} and~\eqref{eq:work_rate}.

\subsection{Definitions and fundamental laws}
We embed the full description [Eq.~\eqref{eq:full_GKSL_eq}] into a closed unitary model~\cite{Esposito2010}. The Markovian bath of the Purcell filter is itself an effective description of unitary dynamics involving additional external modes; enlarging the environment to include these modes, the total state $\rho$ (working medium $+$ environment) obeys the von Neumann equation
\begin{equation}
  \partial_t\rho = -\frac{i}{\hbar}[H(t),\rho],
  \qquad
  H(t) = H_{12}\otimes\mathbf{1}_{\mathrm{env}} + H_{\mathrm{int}} + H_{\mathrm{drive}}(t) + \mathbf{1}_{12}\otimes H_{\mathrm{env}},
\end{equation}
where $H_{12} = \sum_{j=1,2}\hbar\omega_j a_j^\dagger a_j$, $H_{\mathrm{int}} = \sum_{j=1,2}\hbar g_j(a_j^\dagger\sigma_- + a_j\sigma_+)$, and $H_{\mathrm{env}} = (\hbar\omega_q/2)\sigma_z + \hbar\omega_f a_f^\dagger a_f + \hbar g_f(a_f^\dagger\sigma_- + a_f\sigma_+)$. We take a factorized initial state with the environment in the canonical state at the physical inverse temperature $\beta$,
\begin{equation}
  \rho(0) = \rho_{12}(0)\otimes\rho_{\mathrm{env}}^\beta,
  \qquad
  \rho_{\mathrm{env}}^\beta \equiv \frac{e^{-\beta H_{\mathrm{env}}}}{\Tr(e^{-\beta H_{\mathrm{env}}})}.
\end{equation}
The endpoints of the half-cycles of the engine considered in the main text [Eq.~\eqref{eq:endpoints}] approximately satisfy this condition, since the couplings and drive are weak and $e^{\beta_H \hbar\omega_{qH}} \approx e^{\beta_L \hbar\omega_{qL}}$ holds for small $\epsilon$ and large $N$.

Over the interval $[0,T]$ with a single environment, we define the heat delivered to the working medium and the work extracted from it,
\begin{align}
  \label{eq:heat_nonMarkov}
  \mathcal{Q}(T) &\equiv -\int_0^T\! dt\,\Tr\bigl[H_{\mathrm{env}}\,\partial_t\rho(t)\bigr] = -\Delta\langle H_{\mathrm{env}}\rangle,\\
  \label{eq:work_nonMarkov}
  \mathcal{W}(T) &\equiv -\int_0^T\! dt\,\Tr\bigl[(\partial_t H(t))\,\rho(t)\bigr] = -\int_0^T\! dt\,\Tr\bigl[(\partial_t H_{\mathrm{drive}}(t))\,\rho(t)\bigr],
\end{align}
where the second equality of $\mathcal{W}$ uses that only the drive carries explicit time dependence. The first law then takes the form
\begin{align}
  \mathcal{Q}(T) - \mathcal{W}(T)
  &= \int_0^T\! dt\,\Tr\bigl[H_{\mathrm{env}}\,\tfrac{i}{\hbar}[H(t),\rho(t)] + (\partial_t H_{\mathrm{drive}}(t))\rho(t)\bigr]\notag\\
  &= \Delta\langle H_{12} + H_{\mathrm{int}} + H_{\mathrm{drive}}\rangle.
\end{align}
With the working-medium entropy $S_{12}(t) \equiv -\Tr[\rho_{12}(t)\ln\rho_{12}(t)]$, the entropy production over $[0,T]$ is
\begin{equation}
  \Sigma(T) \equiv \Delta S_{12} - \beta\,\mathcal{Q}(T).
\end{equation}
Using that the von Neumann entropy of $\rho$ is conserved under unitary evolution, $\Sigma(T)$ can be rewritten as a relative entropy,
\begin{equation}
  \Sigma(T) = D\bigl(\rho(T)\,\Vert\,\rho_{12}(T)\otimes\rho_{\mathrm{env}}^\beta\bigr) \;\geq\; 0,
\end{equation}
where $D(\rho\Vert\rho') \equiv \Tr[\rho(\ln\rho - \ln\rho')]$, and the last inequality follows from the nonnegativity of relative entropy~\cite{Lindblad1975,Uhlmann1977}.
This is the second law $\Sigma(T)\geq0$.

The corresponding heat current and work rate are
\begin{align}
  \mathcal{J}(t) &\equiv \partial_t\mathcal{Q}(t) = \Tr\bigl[(H_{12} + H_{\mathrm{int}} + H_{\mathrm{drive}}(t))\,\partial_t\rho(t)\bigr],\\
  \dot{\mathcal{W}}(t) &\equiv \partial_t\mathcal{W}(t) = -\Tr\bigl[(\partial_t H_{\mathrm{drive}}(t))\,\rho(t)\bigr],
\end{align}
where $\mathcal{J}$ has been rewritten through the first law. Importantly, neither expression involves any observable of the Markovian bath degrees of freedom, so it is sufficient to work with the density operator on the reduced Hilbert space obtained after tracing out the Markovian bath. Substituting Eq.~\eqref{eq:full_GKSL_eq} into these definitions,
\begin{align}
  \label{eq:exact_mathcal_J}
  \mathcal{J}(t) &= i\hbar\,\Tr\Bigl[\Bigl(-\omega_q\frac{\Omega e^{i\omega_d t}}{2}\sigma_- + g_f\bigl(\textstyle\sum_{j=1,2}g_j a_j^\dagger + \frac{\Omega e^{i\omega_d t}}{2}\bigr)\sigma_z a_f\Bigr)\rho(t)\Bigr] + \mathrm{c.c.}\\
  &\approx -\frac{\omega_q}{\omega_d}\,\Tr\bigl[(\partial_t H_{\mathrm{drive}}(t))\,\rho(t)\bigr],\\
  \label{eq:exact_mathcal_W_dot}
  \dot{\mathcal{W}}(t) &= -\Tr\bigl[(\partial_t H_{\mathrm{drive}}(t))\,\rho(t)\bigr],
\end{align}
where we used $g_f\ll\omega_q$ to drop the second term in $\mathcal{J}$. At this point the ratio between the heat current and the work rate already coincides with that of Eqs.~\eqref{eq:heat_current} and~\eqref{eq:work_rate}.

\subsection{Equivalence with the definitions in the effective description}
To make the equivalence explicit, we pass to the frame of the SWT. We start from the following approximation
\begin{equation}
  \Tr(\partial_t H_{\mathrm{CAS}}(t)\tilde{\rho})=\Tr(\partial_t H_{\mathrm{SWT}}(t)\tilde{\rho}) \approx \Tr(\partial_t H(t) \rho) = \Tr(\partial_t H_{\mathrm{drive}}(t) \rho)
\end{equation}
where we have neglected the total-derivative term $\partial_t\Tr\bigl[\partial_t(e^{S})\,e^{-S}\,\tilde\rho\bigr]$, which contributes at most at order $\lambda$ to the integrated work.
Therefore, within the same approximation, we have
\begin{align}
  \mathcal{J}(t) &\approx -\frac{\omega_q}{\omega_d}\,\Tr\bigl[(\partial_t H_{\mathrm{CAS}}(t))\,\tilde\rho(t)\bigr],\\
  \dot{\mathcal{W}}(t) &\approx -\Tr\bigl[(\partial_t H_{\mathrm{CAS}}(t))\,\tilde\rho(t)\bigr],
\end{align}
where $\tilde\rho$ is the state in the rotating frame defined in Eq.~\eqref{eq:def_of_tilde_rho}. Within the Markov approximation, the integrated work evaluates to
\begin{align}
  \mathcal{W}(T) &= -\int_0^T\! dt\,\Tr\bigl[(\partial_t H_{\mathrm{CAS}}(t))\,\tilde\rho(t)\bigr]\notag\\
  &= i\omega_d\int_0^T\! dt\;e^{-i\omega_d t}\,\Tr\bigl[\hbar g_{\mathrm{CAS}}L_c\sigma_+\,\mathcal{Q}\tilde\rho(t)\bigr] + \mathrm{c.c.}\notag\\
  &\approx i\omega_d\int_0^T\! dt\int_0^t\! ds\;e^{-i\omega_d t}\,
    \Tr\Bigl[\hbar g_{\mathrm{CAS}}L_c\sigma_+\,e^{(\mathcal{L}_{12}+\mathcal{L}_{qf})(t-s)}\,\mathcal{Q}\mathcal{L}_{\mathrm{int}}(s)\bigl(\rho_{12}(s)\otimes\rho_{qf}^{\mathrm{ss}}\bigr)\Bigr] + \mathrm{c.c.}\notag\\
  &\approx \hbar\omega_d\int_0^T\! ds\;\frac{\Gamma_q}{2}g_{\mathrm{CAS}}^2\Bigl(\langle e|\rho_{qf}^{\mathrm{ss}}|e\rangle\,\Tr\bigl[L_c L_c^\dagger\rho_{12}(s)\bigr] - \langle g|\rho_{qf}^{\mathrm{ss}}|g\rangle\,\Tr\bigl[L_c^\dagger L_c\rho_{12}(s)\bigr]\Bigr) + \mathrm{c.c.}\notag\\
  &\approx \frac{\hbar\omega_d}{c}\,\Delta\langle n_2\rangle,
\end{align}
where the third line inserts the leading-order Nakajima--Zwanzig solution for $\mathcal{Q}\tilde\rho$ [Eq.~\eqref{eq:Qrho_integral_rep}] and the fourth uses the bath correlation functions [Eq.~\eqref{eq:Phi_pm_approx}]. 
The last step follows from the identity for the photon transfer induced by the effective CAS dissipator,
\begin{align}
  \label{eq:Dcas_n1}
  \partial_t\langle n_2\rangle &\approx \Tr\bigl(n_2\,(\gamma_{2\to1}\mathcal{D}[L_c] + \gamma_{1\to2}\mathcal{D}[L_c^\dagger])\rho_{12}\bigr)\notag\\
  &\approx c\,\Gamma_q g_{\mathrm{CAS}}^2\Bigl(\langle e|\rho_{qf}^{\mathrm{ss}}|e\rangle\,\Tr[L_c L_c^\dagger\rho_{12}] - \langle g|\rho_{qf}^{\mathrm{ss}}|g\rangle\,\Tr[L_c^\dagger L_c\rho_{12}]\Bigr).
\end{align}
Consequently $\mathcal{Q}(T) \approx (\omega_q/\omega_d)\mathcal{W}(T) \approx (\hbar\omega_q/c) \,\Delta\langle n_2\rangle$, and the unitary definitions reproduce the heat and work obtained from the rates~\eqref{eq:heat_current} and~\eqref{eq:work_rate}, $\mathcal{Q}(T)\approx \int_0^T\,dt J(t)$ and $\mathcal{W}(T)\approx \int_0^T\,dt \dot{W}(t)$.

\section{Analysis of the classical rate equation}
\label{sec:classical_rate_equation}
Here we derive the approximate exponential relaxation of $\langle n_2\rangle$. Equation~\eqref{eq:coherent_GKSL_eq} preserves both the total photon number 
$N = n_1 + n_2$ and the parity of $n_2$. 
Restricting the initial state to a diagonal density matrix supported on 
even values of $n_2$, the dynamics reduces to a classical rate equation 
for the probability distribution $p_{n_2}(t)$:
\begin{align}
  \partial_t p_{n_2}
  &= -(N-n_2+1)(N-n_2+2)
    (n_2-1)n_2 (\gamma_{2\to 1} p_{n_2}-\gamma_{1\to 2}\,p_{n_2-2}) \notag\\
    &+(N-n_2-1)(N-n_2)(n_2+1)(n_2+2) (\gamma_{2\to 1} p_{n_2+2}-\gamma_{1\to 2}\,p_{n_2}).
  \label{eq:rate_equation_wo_approx}
\end{align}
The engine operates near the Gibbs steady state, in which the typical 
$n_2$ is much smaller than the total photon number $N$. 
In this regime the bosonic enhancement factor $n_1(n_1-1)\approx N^2$ 
may be pulled out, and the rate equation reads
\begin{align}
  \partial_t p_{n_2}
  &= -N^2\bigl[
    (n_2-1)n_2\,(\gamma_{2\to 1} p_{n_2}-\gamma_{1\to 2}\,p_{n_2-2}) \notag\\
    &-(n_2+1)(n_2+2)(\gamma_{2\to 1} p_{n_2+2}-\gamma_{1\to 2}\,p_{n_2})
  \bigr].
  \label{eq:rate_eq}
\end{align}
The steady state of Eq.~\eqref{eq:rate_eq} is the Gibbs distribution 
$p_{n_2+2}/p_{n_2} = \gamma_{1\to 2}/\gamma_{2\to 1} \approx e^{-\beta\hbar\omega_q}$, with mean 
$\langle n_2\rangle_\mathrm{ss} \approx 2/(e^{\beta \hbar\omega_{q}}-1)$.
Multiplying Eq.~\eqref{eq:rate_eq} by $n_2$ and summing over even $n_2$ 
gives
\begin{equation}
  \partial_t \langle n_2\rangle = -2N^2\bigl[\gamma_{2\to 1}(\langle n_2^2\rangle - \langle n_2\rangle
  )
  -\gamma_{1\to 2}(\langle n_2^2\rangle + 3\langle n_2\rangle + 2
    )\bigr].
  \label{eq:moment_eq}
\end{equation}
Since the dynamics takes place near the Gibbs steady state, we invoke the 
bosonic fluctuation relation
$\langle n_2^2\rangle 
  \;\approx\; 2\langle n_2\rangle + 2\langle n_2\rangle^2$,
which is exact in the Gibbs state and provides a controlled approximation in 
its vicinity. 
Substituting this into Eq.~\eqref{eq:moment_eq} yields 
\begin{equation}
  \partial_t \langle n_2\rangle
  = -4N^2 (\gamma_{2\to 1}-\gamma_{1\to 2})
    \Bigl(\langle n_2\rangle + 1/2\Bigr)\Bigl(\langle n_2\rangle - \langle n_2 \rangle_{\mathrm{ss}}\Bigr).
  \label{eq:closed_eq}
\end{equation}
On the physical domain $\langle n_2\rangle \geq 0$, 
Eq.~\eqref{eq:closed_eq} therefore describes monotonic relaxation toward 
$\langle n_2\rangle_\mathrm{ss}$. 
Linearizing around $\langle n_2\rangle_\mathrm{ss}$, the deviation 
$\delta n \equiv \langle n_2\rangle - \langle n_2\rangle_\mathrm{ss}$ 
decays exponentially,
  $\partial_t\delta n \;\approx\; -\Gamma\,\delta n$,
  where
  $\Gamma \;=\; 2N^2(\gamma_{2\to 1}+3\gamma_{1\to 2})$.

\section{Numerical details}
\label{sec:numerical_details}

In this section, we provide the numerical details underlying the main result
[Fig.~\ref{fig:performance}] and verify the effective description
[Eq.~\eqref{eq:coherent_GKSL_eq}] against the full description
[Eq.~\eqref{eq:full_GKSL_eq}].

\subsection{Numerical method}

In the frame rotating at the drive frequency $\omega_d$,
Eq.~\eqref{eq:full_GKSL_eq} reduces to the time-independent GKSL equation
\begin{align}
  \partial_t\rho
  = -\frac{i}{\hbar}[H_R,\rho]
    + \kappa(n_\text{th}+1)\,\mathcal{D}[a_f]\rho
    + \kappa\, n_\text{th}\,\mathcal{D}[a_f^\dagger]\rho
  \equiv \mathcal{L}\rho,
  \label{eq:rotating_frame_GKSL}
\end{align}
with the static Hamiltonian
$H_R = \sum_{j=1,2,f}\bigl[\hbar(\omega_j-\omega_d)\,a_j^\dagger a_j
+ \hbar g_j\bigl(a_j^\dagger\sigma_- + a_j\sigma_+\bigr)\bigr]
+ \hbar(\omega_q-\omega_d)\,\sigma_z/2 + (\hbar\Omega/2)\,\sigma_x$.
The dynamics can therefore be obtained from the spectral decomposition of the
Liouvillian $\mathcal{L}$, rather than by direct time integration of the
time-dependent master equation [Eq.~\eqref{eq:full_GKSL_eq}].
Since full diagonalization of $\mathcal{L}$ becomes numerically intractable
for large total photon number $N$, we consistently employ the shift-invert
method to extract the $500$ slowest-decaying eigenmodes of $\mathcal{L}$,
i.e., those with eigenvalues closest to zero, which govern the long-time
dynamics relevant to the engine operation; all data in
Fig.~\ref{fig:performance} are obtained in this way.
The local Hilbert spaces are truncated at the photon-number cutoffs
$n_{1,\max}=N+2$, $n_{2,\max}=2$, and $n_{f,\max}=2$, which are sufficient in
the low-temperature regime in which the engine operates.

The initial state, defined in the laboratory frame, is taken to be
\begin{align}
  \rho(0) &= (1-\delta)\,\rho_L + \delta\,\rho_H,
  \nonumber\\
  \rho_\nu &\propto
  U_\nu^\dagger
  \Biggl[
    \Biggl(\;
      \sum_{n_2=0,2,\dots,2\lfloor N/2\rfloor}
      e^{-\beta_{\mathrm{eff},\nu}\, c\hbar(\omega_2-\omega_1)\, n_2}\,
      \ketbra{N-n_2,\,n_2}{N-n_2,\,n_2}
    \Biggr)
    \otimes \rho^{\mathrm{ss}}_{qf,\nu}
  \Biggr]
  U_\nu,
  \label{eq:initial_state}
\end{align}
where $U_\nu$ is the SWT unitary of stroke $\nu$ evaluated at $t=0$.
This choice places the engine close to the lower endpoint $n_{2-}$ of the
steady cycle [cf.\ Eq.~\eqref{eq:endpoints}], and the state is supported on
the superselection sector labeled by the total photon number $N$ and the even
parity of $n_2$.

\subsection{Numerical verification of the effective description}

\begin{figure*}
  \includegraphics[width=\linewidth]{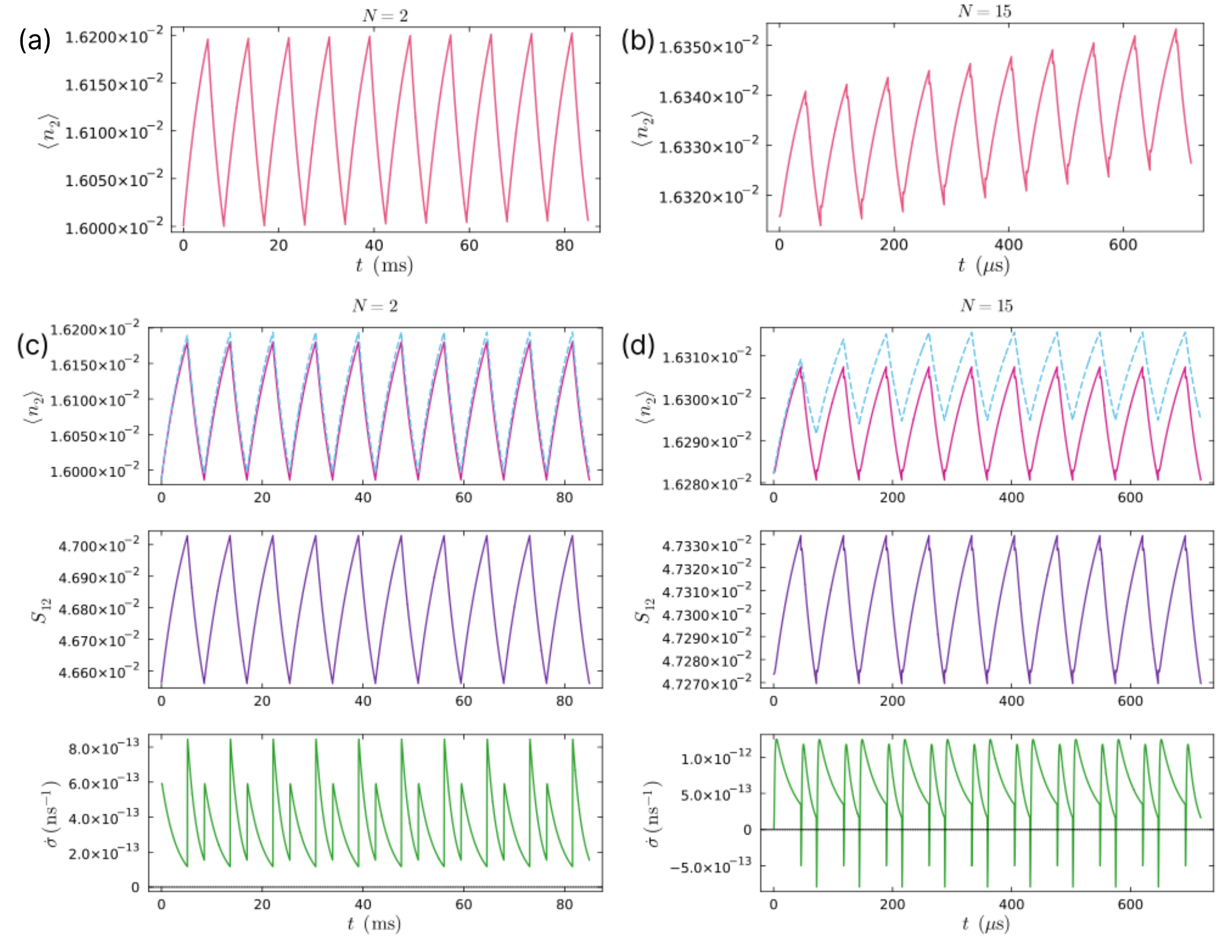}
  \caption{Numerical verification of the effective description.
  (a),(b)~First ten cycles of $\langle n_2\rangle$ obtained from the full
  master equation [Eq.~\eqref{eq:full_GKSL_eq}] without projection, for
  $N=2$~(a) and $N=15$~(b); the slow drift visible for large $N$ originates
  from relaxation between different superselection sectors.
  (c),(d)~Dynamics of $\langle n_2\rangle$ (top), the von Neumann entropy
  $S_{12}$ (middle), and the entropy production rate $\dot\sigma$ (bottom),
  evaluated with the density matrix projected onto the initial
  superselection sector, for $N=2$~(c) and $N=15$~(d).
  Solid magenta and dashed light-blue curves show the full
  [Eq.~\eqref{eq:full_GKSL_eq}] and effective
  [Eq.~\eqref{eq:coherent_GKSL_eq}] dynamics, respectively.
  The entropy production rate is defined as
  $\dot\sigma \equiv \partial_t S_{12}
  - \beta_\nu (\hbar\omega_{q\nu}/2)\,\partial_t\langle n_2\rangle$
  [cf.\ Eq.~\eqref{eq:JW_compact} with $c=2$], and is evaluated by
  numerically differentiating the computed dynamics of $S_{12}$ and
  $\langle n_2\rangle$.
  After the projection, both $\langle n_2\rangle$ and $S_{12}$ return to
  their initial values once per cycle, and $\dot\sigma$ remains nonnegative
  throughout the strokes.
  For $N=15$, $\dot\sigma$ takes negative values at the bath-switching
  instants; this is a transient effect caused by the momentary breakdown of
  the Markov approximation and is negligible in the thermodynamic evaluation
  of the cycle.}
  \label{fig:sm_verification}
\end{figure*}

Figures~\ref{fig:sm_verification}(a) and \ref{fig:sm_verification}(b) show
the first ten cycles of the engine dynamics generated by the full GKSL
equation [Eq.~\eqref{eq:full_GKSL_eq}] for $N=2$ and $N=15$, respectively.
The performance of a heat engine must be assessed over a closed cycle, in
terms of both the internal energy
$E = \hbar(\omega_2-\omega_1)\langle n_2\rangle + \mathrm{const.}$
and the entropy $S_{12}$.
For large $N$, however, very slow relaxation modes of the full Liouvillian
that connect different superselection sectors become visible on the timescale
of the simulation.
These modes cause a gradual drift of both $\langle n_2\rangle$ and $S_{12}$
and thereby prevent the cycle from closing.

To remove these spurious contributions, we evaluate all physical quantities
with the density matrix projected onto the initial superselection sector and
subsequently renormalized.
Figures~\ref{fig:sm_verification}(c) and \ref{fig:sm_verification}(d) show
the resulting dynamics of $\langle n_2\rangle$, the von Neumann entropy
$S_{12}$, and the entropy production rate $\dot\sigma$ for $N=2$ and $N=15$,
respectively.
The power reported in the main text [Fig.~\ref{fig:performance}] is likewise
computed from the change of $\langle n_2\rangle$ of the projected density
matrix.
For $N=2$, the full and effective dynamics nearly coincide, whereas for
$N=15$ a discrepancy develops between the two descriptions; this discrepancy
corresponds to the superlinear growth of the power visible in Fig.~\ref{fig:performance}.
We find that both $\langle n_2\rangle$ and $S_{12}$ return to their initial
values after each cycle, even in the large-photon-number regime.
Hence, once the slow leakage between superselection sectors is projected out,
the engine operation is thermodynamically consistent: the cycle closes in
both energy and entropy, and the entropy production rate remains nonnegative
throughout the strokes.

\end{document}